\documentclass[prx,10pt,a4paper,twocolumn,nofootinbib, superscriptaddress]{revtex4-1}

\expandafter\let\csname equation*\endcsname\relax
\expandafter\let\csname endequation*\endcsname\relax
\expandafter\let\csname eqnarray*\endcsname\relax
\expandafter\let\csname endeqnarray*\endcsname\relax

\usepackage{amsmath,mathtools}
\usepackage{amssymb}	
\usepackage{graphicx}
\usepackage{braket}
\usepackage{color}
\usepackage{float}
\usepackage{soul}
\usepackage[hidelinks]{hyperref}

\newcommand{\A}{\nonumber}
\newcommand{\RN}[1]{\textup{\uppercase\expandafter{\romannumeral#1}}}

\begin{document}
\setlength{\parindent}{0mm}

%======================================================================================
\title{Rotation of polarization in the gravitational field of a laser
  beam -  Faraday effect and optical activity} 
\author{Fabienne Schneiter}
\email{fabienne.schneiter@uni-tuebingen.de }
\address{Eberhard-Karls-Universit\"at T\"ubingen, Institut f\"ur Theoretische Physik, 72076 T\"ubingen, Germany}

\author{Dennis R\"atzel}
\email{dennis.raetzel@physik.hu-berlin.de}
\affiliation{Institut f\"ur Physik, Humboldt-Universit\"at zu Berlin, Newtonstraße 15, 12489 Berlin, Germany}

\author{Daniel Braun}
\address{Eberhard-Karls-Universit\"at T\"ubingen, Institut f\"ur Theoretische Physik, 72076 T\"ubingen, Germany}

%======================================================================================
\begin{abstract}{\noindent
We investigate the rotation of the polarization of a light ray propagating in the gravitational field of a circularly polarized laser beam. The rotation consists of a reciprocal part due to gravitational optical activity, and a non-reciprocal part due to the gravitational Faraday effect. We discuss how to distinguish the two effects: Letting light propagate back and forth between two mirrors, the rotation due to gravitational optical activity cancels while the rotation due to the gravitational Faraday effect accumulates. In contrast, the rotation due to both effects accumulates in a ring cavity and a situation can be created in which gravitational optical activity dominates. Such setups amplify the effects by up to five orders of magnitude, which however is not enough to make them measurable with state of the art technology. The effects are of conceptual interest as they reveal gravitational spin-spin coupling in the realm of classical general relativity, a phenomenon which occurs in perturbative quantum gravity.}
\end{abstract}

%======================================================================================
\maketitle

%======================================================================================
\section{Introduction}

The gravitational field of a light beam was first studied in 1931 by
Tolman, Ehrenfest and Podolski \cite{tolman}, who described the laser
beam in the simplest way, namely as a single light ray of constant
energy density and without polarization. Since then, various models
for light beams have been considered, such as in
\cite{bonnor_1969,raetzel_2016, strohaber_2018}, all of them having in
common that the short-wavelength approximation is used. This means
that the light is either described as a thin pencil or as a continuous
fluid moving at the speed of light and without any wave-like
properties. Recently, we studied the gravitational field of a laser
beam beyond the short-wavelength approximation \cite{gaussstrahl}: The
laser beam is modeled as a solution of Maxwell's equations, and
therefore, has wave-like properties. In this case, there appear
gravitational effects of light that were not visible in the previous
models, such as frame-dragging due to the light's spin
angular-momentum,
 the deflection of a parallel propagating test ray, and
the rotation of polarization of test rays. The latter is the
subject of this article. 

Effects of gravitational rotation of polarization were first described in 1957 independently by Skrotsky \cite{skrotsky_1957} and by Balazs \cite{balazs_1957}.
In 1960, Plebanski found a coordinate-invariant expression for the
change of the polarization for a light ray coming from flat spacetime,
passing through a weak gravitational field, and going to flat
spacetime again \cite{plebanski_1960}. 
The gravitational rotation of polarization has been
studied for several systems: for moving objects, moving gravitational
lenses \cite{lyutikov_2017, kopeikin_2002, pen} and other
astrophysical situations \cite{sereno_2004,chen_2015}, in the context
of gravitational waves \cite{piran_1985}, for rotating rings
\cite{dehnen_1972}, for ring lasers \cite{cox_2006} and for linearly
polarized lasers in a waveguide \cite{Ji:2006grav}. It was also
treated more formally in \cite{gnedin_1988,produtch_2011,brodutch_2011}.

Rotation of polarization is well-known from classical optics, when light rays pass through certain media (see e.g.\cite{LL}). For this, the medium needs broken inversion symmetry, a property certain materials have naturally. Such  media with ``natural optical activity'' lead to different phase velocities of right- and left-circularly polarized light.  The effect is ``reciprocal'', i.e.~when the light ray is reflected back through the medium, the rotation of polarization is undone.  In contrast hereto is the Faraday effect, which can be created even in isotropic media by applying a magnetic field. Here, the rotation is ``non-reciprocal'', i.e.~the polarization keeps rotating in the same direction relative to the original frame when the light propagates back along the path.     
In this article, we consider the rotation of the polarization vector of test rays in the gravitational field of a circularly polarized laser beam in free space.  
It turns out that the rotation of polarization 
contains both a reciprocal and a non-reciprocal part.
The former can hence be interpreted as 
gravitational optical activity 
and the latter as a gravitational Faraday effect,
also called Skrotsky effect.

The laser beam is described as a perturbative solution to Maxwell's equations, an expansion in the beam divergence angle $\theta$, which is assumed to be smaller than one radian. The description of the laser beam and its gravitational field is given in detail in \cite{gaussstrahl} and summarized below. 
We look at the rotation of the polarization of test rays which
are parallel co-propagating, parallel counter-propagating, or
propagating transversally to the beamline of the source
laser-beam, and consider a cavity where the rotation of the polarization vector accumulates after each roundtrip. { We thus propose a measurement scheme which may possibly be realized in a laboratory in the future, when the sensitivity in experiments has improved accordingly.}

The description of the gravitational field of a laser beam is reviewed in section~\ref{gauss}, and the calculation
of the rotation of light polarization in curved spacetime in section~\ref{rot}. In
section~\ref{ray}, we calculate the Faraday effect for test
rays. As already mentioned, only the non-reciprocal part of the
rotation which is not due to the deflection can be associated with the
Faraday effect, which is discussed in section~\ref{Ff}. Considering a
cavity in a certain arrangement, the rotation angles acquired after
each roundtrip of the light inside the cavity sum up. This is the
subject of section~\ref{cavity}, where we look at a one-dimensional
cavity and a ring cavity and discuss the possible measurement
precision of the rotation angle. We give a conclusion and an outlook in
section~\ref{conc}. 

To keep track of the orders of magnitude, we introduce dimensionless
coordinates by dividing them by the beam waist $w_0$ as $\tau=ct/w_0$,
$\xi=x/w_0$, $\chi=y/w_0$ and $\zeta=z/w_0${, where $c$ is the speed of light}. 
Greek indices like
$x^\alpha$ refer to dimensionless spacetime coordinates and
latin indices like $x^a$ refer to dimensionless spatial
coordinates. For the spacetime metric, we choose the sign convention
$(-,+,+,+)$, such that the Minkowski metric $\eta$ in the dimensionless coordinates reads
$\eta=w_0^2\mathrm{diag}(-1,1,1,1)$. In the
numerical examples and plots, we use the following values: the beam waist
$w_0=10^{-6}\,{\rm m}$, the beam divergence $\theta=0.3\,$rad (this
determines the wavelength, which is given by $\pi\theta w_0\simeq
1\,{\rm \mu m}$), the polarization $\lambda=1$, and the power of the
source laser-beam, which is directed in the positive $z$-direction,
$P_0=10^{15}\,{\rm W}$.

%======================================================================================
\section{The gravitational field of a laser beam beyond the short wavelength approximation \label{gauss}} 

\begin{figure}[H]\center
\includegraphics[scale=0.8]{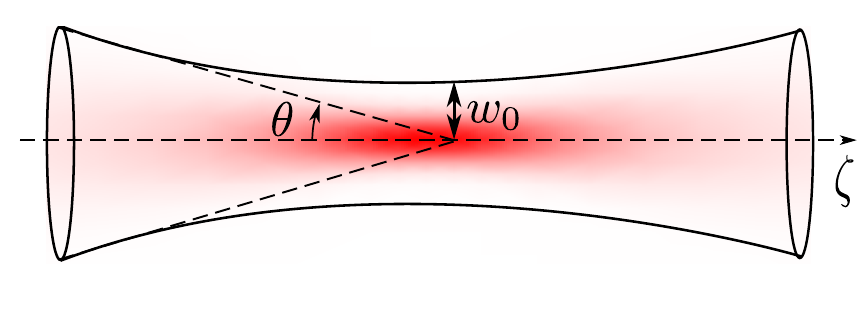}
\caption{\label{beam}Schematic illustration of the laser beam
  propagating in the positive $\zeta$-direction: The beam divergence
  $\theta$ describes the opening angle of the laser beam and is
  assumed to be a small parameter (smaller than $1\,{\rm rad}$), and
  the beam waist $w_0$ is a measure for the radius of the laser beam
  at its focal point. The intensity of the laser beam decreases with a
  Gaussian factor with the distance from the beamline.} 
\end{figure}

In this section, we summarize the description of the laser beam and its
gravitational field presented in \cite{gaussstrahl}. 
A laser beam is accurately described by a Gaussian beam. The Gaussian
beam is a monochromatic electromagnetic, almost plane wave whose
intensity distribution decays with a Gaussian factor with the distance
to the beamline. It is obtained as a perturbative solution of
Maxwell's equations, namely an expansion in the beam divergence $\theta$, the
opening angle of the beam, which is assumed to be smaller than one
radian.  
The electromagnetic four-vector potential describing the Gaussian beam
is obtained by a plane wave multiplied by an envelope function that
is assumed to vary slowly in the direction of propagation, in
agreement with the property that the divergence of the beam is
small. Corresponding to these features, one makes the ansatz for the
four-vector potential $A_\alpha(\tau,\xi,\chi,\zeta)=\tilde{\mathcal{A}}
v_\alpha(\xi,\chi,\theta \zeta)e^{i\frac{2}{\theta }(\zeta-\tau)}$,
where $\tilde{\mathcal{A}}$ is the amplitude and $v_\alpha$ the
  envelope function.\footnote{More precisely, the complex-valued
  vector potential $A$ we consider is the analytical signal of the
  real-valued physical vector potential, which is obtained by taking
  the real part of $A$.} The exponential factor describes a plane wave
propagating in $\zeta$-direction with wave number $k={2}/({\theta
  w_0})$, where $w_0$ is the beam waist at its focal point. The laser
beam propagates in positive $\zeta$-direction such that its beamline
concides with the $\zeta$-axis. The beam is illustrated in
figure~\ref{beam}. 
Like the four-vector potential for any radiation, $A_\alpha$ satisfies
the Maxwell equations, which, in vacuum, are given by the wave equations
\begin{equation}\label{eq:maxwellfour}
	\left(-\partial_\tau^2+\partial_\xi^2+\partial_\chi^2
          +\partial_\zeta^2\right) 
A_\mu(\tau,\xi,\chi,\zeta)=0\,,
\end{equation}
where the Lorenz-gauge condition is chosen. 
Since the envelope function varies slowly in the direction of
propagation, the wave equations (\ref{eq:maxwellfour}) reduce to a
Helmholtz equation for each component of the envelope function,  
\begin{equation}
	\left(\partial_\xi^2+\partial_\chi^2+\theta^2 \partial_{\theta \zeta}^2+4iw_0\partial_{\theta \zeta}\right) v_{\alpha}(\xi,\chi,\theta \zeta)=0\,.
\end{equation} 
This Helmholtz equation is solved by writing the envelope function as
a power series in the small parameter $\theta$. 
One obtains an equation for each order of the expansion of the
envelope function, with a source term consisting of the solution for a
lower order, where even and odd orders do not mix. 
The beam is assumed to have left-~or right-handed circular
polarization, which we label by $\lambda=\pm 1$.\footnote{The vector
  potential describing the laser beam thus depends on the parameter
  $\lambda$, and so do its energy-momentum tensor, the induced metric
  perturbation and the effects we calculate in the following
  sections. Therefore, these quantities can be thought of as
    being labelled by an index $\lambda$, which we suppress in the
  following, except for appendix~\ref{A}, where we write the index $\lambda$ explicitly.} We define this to be the case if its field strength, defined as
$F_{\alpha\beta}=\partial_\alpha A_\beta-\partial_\beta A_\alpha$, is
an eigenfunction with eigenvalue $\pm 1$ of the generator of the
duality transformation of the electromagnetic  field given by
$F_{\alpha\beta}\mapsto -i
\epsilon_{\alpha\beta\gamma\delta}F^{\gamma\delta}/2$, where
$\epsilon_{\alpha\beta\gamma\delta}$ is the completely anti-symmetric
tensor with $\epsilon_{0123}=-1$. Our definition of helicity is based on the invariance of Maxwell's equations under the duality transformation and the conservation of the difference between photon numbers of right- and left-polarized photons shown in \cite{calkin_1965} (see also
\cite{Trueba1996the,Barnett2012dup,Fernandez-Corbaton2012hel,Andersson:2018juv}). 
For $\theta=0$,
this leads to the usual expressions for the field strength of a
circularly polarized laser beam. 

It turns out that the energy-momentum tensor, which one may expect to
be oscillating at the frequency of the laser beam, does not contain
any oscillating terms when circular polarization is assumed. The
energy-momentum tensor reads (see appendix A for the explicit
expressions) 
\begin{equation}
	T_{\alpha\beta}=\frac{c^2\epsilon_0}{2}\;{\rm
          Re}\left(F_{\alpha}^{\;\sigma}F^*_{\beta\sigma}-\frac{1}{4}\eta_{\alpha\beta}F^{\delta\rho}F^*_{\delta\rho}\right)\,. 
\end{equation}
The power series expansion of the envelope function induces a
  power series expansion of the energy-momentum tensor and the
  expansion coefficients are identified as different order terms of
  $T_{\alpha\beta}$ in $\theta$.

Since the energy density of a laser beam is small compared to the
  one of ordinary matter, one may expect its gravitational field to
be weak. The spacetime metric describing the gravitational field is
thus assumed to consist of the metric for flat spacetime
$\eta_{\alpha\beta}$ plus a small perturbation
$h_{\alpha\beta}$. Terms quadratic in the metric perturbation are
neglected; this is the linearized theory of general relativity. In
this case, the Einstein field equations reduce to wave equations for
the metric perturbation \cite{mtw} 
\begin{equation}\label{eq:einsteinlin}
	\frac{1}{w_0^2}\left(-\partial_\tau^2+\partial_\xi^2+\partial_\chi^2+\partial_\zeta^2\right) h_{\alpha\beta} = -\frac{16\pi G}{c^4} T_{\alpha\beta}\,,
\end{equation} 
where $G$ is Newton's constant and where the Lorenz-gauge has been chosen. 
Like the envelope function and the energy-momentum tensor, the metric perturbation is expanded in the beam divergence, \begin{equation}
	h_{\alpha\beta}(\xi,\chi,\theta \zeta)= \sum_{n=0}^\infty
        \theta^n h_{\alpha\beta}^{(n)}(\xi,\chi,\theta \zeta)\,. 
\end{equation} 
For a laser beam extending from minus to plus spatial infinity,
the wave equations (\ref{eq:einsteinlin}) result in a two-dimensional Poisson equation
for each $h_{\alpha\beta}^{(n)}$, with a source term consisting
  of a term of the energy-momentum tensor of the same order and a term
  proportional to $h_{\alpha\beta}^{(n-2)}$, where even and odd
orders do not mix. Details and the solutions for the zeroth, the
first and the third order, which are relevant for our purposes, are
given in appendix~\ref{A}. 

For a finitely extended source beam, the solution of \eqref{eq:einsteinlin}
with time-independent
energy-momentum tensor of the source laser-beam can be calculated
using the Green's function of the three-dimensional Poisson
equation, 
\begin{eqnarray}\label{eq:primeli}
	 h_{\alpha\beta}
	&=&	\frac{4Gw_0^2}{c^4}\int d\xi' d\chi' d\zeta' 
		\frac{T_{\alpha\beta}\left(\xi',\chi',\theta\zeta'\right)}{|\vec{x}-\vec{x}'|}\;,
\end{eqnarray}
where $\vec{x}=(\xi,\chi,\zeta)$ and $\vec{x}'=(\xi',\chi',\zeta')$.
The solution (\ref{eq:primeli}) is discussed in detail in
\cite{gaussstrahl}.

%======================================================================================
\section{Rotation of polarization in a weakly curved spacetime\label{rot}}

In this section, we explain the expression presented in
\cite{plebanski_1960} for the rotation angle that
the polarization vector
of a test ray
acquires when propagating in a gravitational
field. 

For a light ray propagating through a gravitational field and starting
and ending at spatial infinity, the rotation angle of polarization
within a plane perpendicular to the propagation direction (in the
following called ray-transverse plane) is given by  equation (5.33) in
\cite{plebanski_1960}. For our set of coordinates, it takes the form 
\begin{equation}
\Delta
= \frac{1}{2w_0^2} \int_{-\infty}^\infty d \tau\;
 t_0^a \epsilon_{abc} \partial_b h_{c\alpha}(\tau,\varrho_\perp + \tau t_0 ) t_0^\alpha\;,
\label{eq:rotpleb}
\end{equation} 
where $\epsilon_{abc}$ is the Levi-Civita symbol in three dimensions
with $\epsilon_{123}=1$, $t^a_0=\dot\gamma^a(\tau_0)$ is the initial
tangent to the curve describing the light ray parametrized by the
dimensionless parameter $\tau$, and the line $\varrho_\perp + \tau
t_0$ with $\varrho_\perp=(\xi_0,\chi_0,0)$ constant is equivalent to the
spatial part of the ray $\gamma$ 
including terms up to linear order in the metric
perturbation. Therefore, the evaluation of the metric perturbation along the line
$\varrho_\perp + \tau t_0$ instead of {$\gamma$ {the actual, possibly deflected
trajectory of a light ray in the gravitational field of the source}}
is justified as the correction would be of higher order in the
metric perturbation.

The sign of the rotation { angle $\Delta$} is chosen such that the positive sign
refers to right-handedness (handedness of rotation as inferred
from equation (5.20) of \cite{plebanski_1960}). Equation (\ref{eq:rotpleb}) was obtained
using the formal analogy of Maxwell's equations in a dielectric medium
and Maxwell's equations in a gravitational field and using geometric
ray optics for vectors. It is shown in \cite{plebanski_1960} that the
expression in equation (\ref{eq:rotpleb}) is invariant under coordinate
transformations that approach the identity at spatial infinity.
For equation (\ref{eq:rotpleb}) to be valid, the metric perturbation and all its
first derivatives have to vanish at least as $\tilde\rho^{-1}$ for
$\tilde\rho\rightarrow \infty$, where
$\tilde\rho=\sqrt{\xi^2+\chi^2+\zeta^2}$. 

\begin{figure}[H]\center
\includegraphics[scale=0.8]{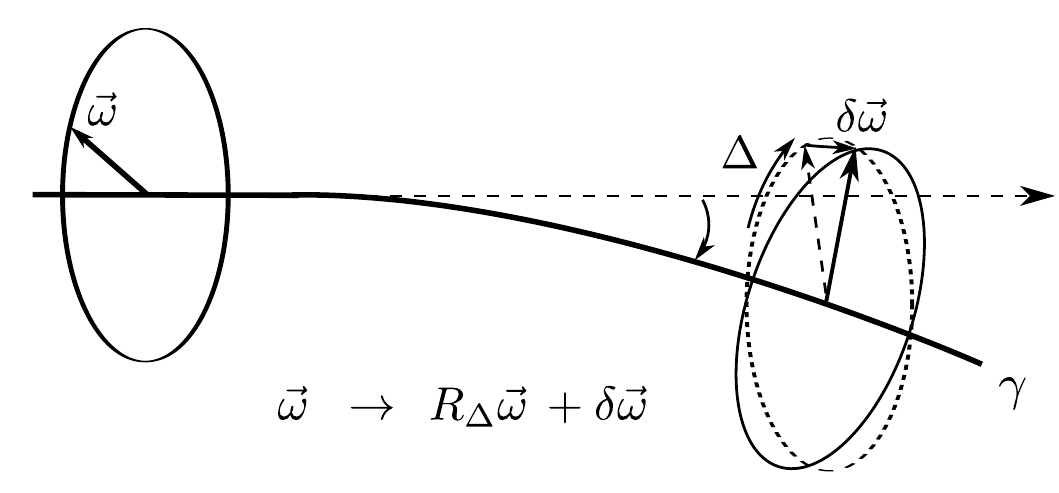}
\caption{Change of the initial polarization vector $\vec\omega$ of a
  light ray $\gamma$: The initial polarization vector $\vec\omega$ in
  the initial ray-transverse plane (represented by the solid
  circle on the left and the dashed circle on the right) is rotated by
  the angle $\Delta$ into $R_\Delta\vec\omega$ (dashed arrow on the
  right) due to the gravitational field, where $R_\Delta$ is the
  corresponding rotation matrix. Additionally, the deflection of the
  laser beam tilts the ray-transverse plane into its final orientation
  (solid circle on the right) such that it stays orthogonal to the
  tangent of the deflected laser beam. The tilt leads to an additional
  change $\delta\vec\omega$ of the polarization vector
  $\vec\omega$. The rotation by the angle $\Delta$ consists of a
  reciprocal part due to the  gravitational optical
  activity and a non-reciprocal part due to the gravitational Faraday
  effect. } 
\label{figrot}
\end{figure}

For a light ray that is not deflected by the gravitational field,
i.e.~that does not change its direction of propagation, the
  ray-transverse plane is the same everywhere far away from the laser
beam, where spacetime is flat. However, when the light ray is
deflected, this plane is tilted after passing the gravitational field
with respect to the one before entering the gravitational
field. Therefore, the rotation of the polarization vector within the
ray-transverse plane given in equation (\ref{eq:rotpleb}) is
superimposed with a change of the polarization vector
$\delta\vec\omega$ due to the deflection of the light ray. The latter
consists of a rotation plus a deformation which depend on the initial
polarization vector $\vec\omega$.\footnote{See section~6 in
  \cite{plebanski_1960}.}  
It does not contribute to the gravitational Faraday effect or { the}  optical activity. An experimentalist
who wants to measure  these effect would thus have to correct for  the
deflection effects. 
The change of the polarization vector is illustrated in figure~\ref{figrot}.
Another approach how to describe the rotation of polarization is
described in appendix~\ref{B}. It agrees with the results presented in this
section. 

For a linearly polarized test ray, the interpretation of the
rotation of the polarization vector is clear: For example, { for a
  test light-ray propagating { in} $\zeta$-direction,} the 
polarization vector describing linear polarization in $\xi$-direction,
$\vec\epsilon_\xi=(1,0,0)$, is rotated into $R_\Delta\vec
\epsilon_\xi=(\cos(\Delta),\sin(\Delta),0)$, where $R_\Delta$ stands
for the matrix rotating by the angle $\Delta$ about the
$\zeta$-axis. For a circularly polarized test ray with helicity
$\lambda_{\rm test}=\pm 1$ and with the corresponding polarization
vector $\vec\epsilon_{\lambda_{\rm test}} =
(1,-\lambda_{\rm test} i,0)/\sqrt{2}$, one obtains
$R_\Delta\vec\epsilon_{\lambda_{\rm test}} = e^{i\lambda_{\rm test}
  \Delta} \vec\epsilon_{\lambda_{\rm test}} $. This means that the circularly
polarized test ray acquires the phase $\lambda_{\rm
  test}\Delta$.
In general, for an elliptically polarized test
light ray, the acquired phases of the circular components 
lead to a
rotation of the major axis of the ellipse by an angle $\Delta$.

%======================================================================================
\section{Rotation of polarization in the gravitational field of a laser beam}\label{ray}

In this section, we investigate the rotation of the polarization
vector of a test ray passing through the gravitational field 
of a source laser-beam according to equation (\ref{eq:rotpleb}).

We consider different orientations of the test ray with
respect to the source beam: parallel co-propagating, parallel
  counter-propagating,
and transversal test rays. We find that the effect depends strongly
on the orientation of the test ray. In particular, we
obtain that the order of the metric
expansion\footnote{Generally, with the order of the metric expansion,
  we refer to the order in $\theta$. Any higher order terms of the
  metric perturbation itself are neglected in the linearized theory of
  general relativity.} that causes the rotation of polarization
depends on the orientation of the test ray.

The source laser-beam is assumed to propagate along the $\zeta$-axis,
to be emitted at $\zeta=\alpha$ and absorbed at $\zeta=\beta$. The
{parallel co-propagating test ray is emitted at $\zeta=A$ and absorbed
  at $\zeta=B$ and the parallel counter-propagating test ray is
  emitted at $\zeta=B$ and absorbed at $\zeta=A$. The test ray that is
  oriented transversally to the beamline of the source laser-beam is
  emitted at $\xi=A$ and absorbed at $\xi=B$ { or vice versa}.}  

In subsection~\ref{inf} we focus on an ideal situation of
infinitely long test rays. The source laser-beam is considered to be either finitely or infinitely extended. In subsection~\ref{finite} we look at finitely long test rays and a finitely extended source laser-beam, and we discuss the the long-range behavior of the rotation of polarization of the test rays. In subsection~\ref{spin}, we discuss the gravitational coupling between the spin of the source laser-beam and the spin of the test ray.

%======================================================================================
\subsection{Infinitely extended test ray\label{inf}}

For the infinitely extended test rays, the conditions for the application of equation (\ref{eq:rotpleb}) are
immediately seen to be fulfilled  for the finitely
extended source beam, as the metric perturbation and all its first derivatives
vanish at least as
$\tilde\rho^{-1}$ for $\tilde\rho\rightarrow\infty$.  This follows
directly from the Green's function which is proportional to $1/\tilde{\rho}$ in three dimensions.
  
For the parallel test rays, for an infinitely extended
source beam and an infinitely extended test ray
it will always be understood that the
emitter and 
absorber of the test ray are sent to infinity more rapidly than
those of the source-beam, i.e.~
$|A|,|B|\gg |\alpha|,|\beta| \to\infty$, such that also here the test ray indeed
begins and ends in flat spacetime.    
For the transversal test rays, for an infinitely extended source beam
and infinitely extended test rays, it is assumed that $|A|$ and $|B|$
approach infinity fast enough for them to be in flat spacetime.  

Besides the strict validity of equation~\eqref{eq:rotpleb}, the infinite
test-ray has also the advantage to lead to relatively simple
analytical expressions for the rotation angles.

%-----------------------------------------------------
\subsubsection{Parallel test rays\label{parall}}

\begin{figure}[H]\center
\includegraphics[scale=0.8]{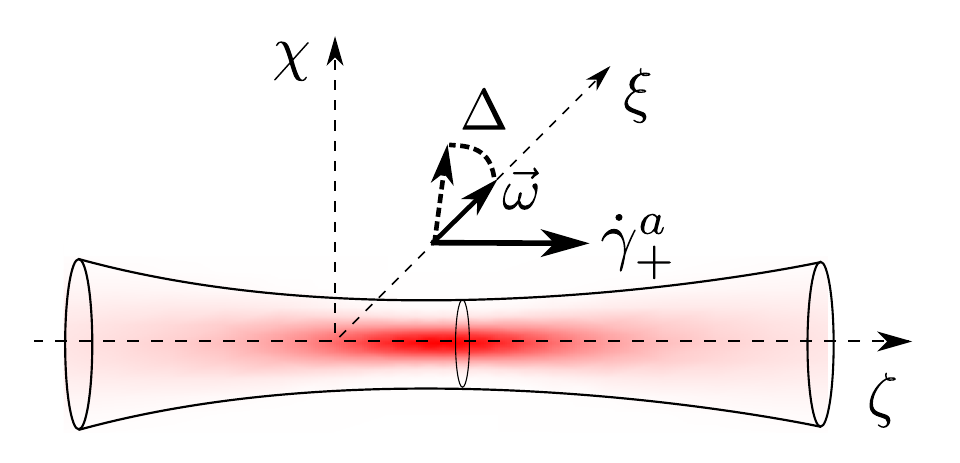}
\caption{\label{33}Schematic illustration of the rotation of the polarization vector $\vec\omega$ (here it originally has only a
component in $\xi$-direction) of a parallel co-propagating test ray with tangent $\dot\gamma_+$ in the gravitational field of
the laser beam.} 
\end{figure}

We start by looking at the rotation of the polarization vector of test rays which are parallel co-propagating or counter-propagating
with respect to the source laser-beam as illustrated in
figure~\ref{33}. 
The parallel co-~and counter-propagating test rays are assumed
to have a distance $\rho=\sqrt{\xi^2+\chi^2}$ from the beamline, 
and to
travel from $\zeta=-\infty$ to $\zeta=\infty$ and from $\zeta=\infty$
to $\zeta=-\infty$, respectively. They are considered to have
transversal polarization described by the polarization vector
$w^\mu=(0,w^\xi,w^\chi,0)$.  
The initial tangents to their worldlines are given by
${ \dot\gamma_\pm}(\tau_0)=(1,0,0,\pm(1-f^\pm))$, where the "$+$"
corresponds to the co-propagating test ray and the "$-$" to the
counter-propagating test ray. The parameter $f^\pm$ ensures that
{ $\dot\gamma_\pm$} satisfies the null condition. It is proportional
to the metric perturbation, which means that it does not contribute in
equation (\ref{eq:rotpleb}) and can be neglected in the following
calculations.   
Since the integration in equation (\ref{eq:rotpleb}) is along the line
$\varrho_\perp + \tau t_0 = (\xi_0,\chi_0,\pm\tau)$, we can
change the integration variable from $\tau$ to $\zeta$ when neglecting
terms quadratic in the metric perturbation. 
Then, for the parallel propagating test rays we obtain (see
equation (\ref{pmapp})) 
\begin{align}\label{pm} 
\Delta_{\pm}
= -\frac{1}{2w_0^2}\int_{-\infty}^\infty { d\zeta} &
 \Big(\partial_\chi\left(h_{\xi\zeta} \pm h_{\tau\xi}\right)  -\partial_\xi\left(h_{\chi\zeta} \pm h_{\tau\chi}\right)\Big)\,.
\end{align}
Notice that the metric perturbation contains a factor $w_0^2$, such that $\Delta_{\pm}$ is dimensionless. 
For the co-propagating test ray, the contribution {coming from
  the first order of the metric perturbation} cancels, and one obtains
in leading order (the third order in $\theta$)
\begin{equation}\label{bla}
\Delta_+
= { \lambda\frac{GP_0\theta^3}{c^5} \int_{\alpha}^\beta
d\zeta\,}
|\mu|^2(1+2|\mu|^2\rho^2)
e^{-2|\mu|^2\rho^2}\;,
\end{equation}
where $|\mu|^2=(1+(\theta\zeta)^2)^{-1}$. Note that $\zeta$ in
  \eqref{bla} parametrizes the source beam (i.e.~corresponds to
  $\zeta'$ in \eqref{eq:primeli}).  The derivation of
  \eqref{bla} (see appendix~\ref{fininf} for details) uses an asymptotic
  expansion in $1/B$, i.e.~assumes 
  that $B\gg |\zeta'|,|\rho'|$, as well as a finite cut-off $\rho_0$ of the
  energy density in radial direction that is then sent to infinity.
  The expression with $\rho_0$ kept finite is given by
  \eqref{blafin}. 
For an infinitely extended source beam, we can then simply evaluate
the limit $\alpha\rightarrow -\infty$ and $\beta\rightarrow \infty$. 
{ An alternative derivation that starts from an infinitely extended
source beam and an infinitely extended test ray is given in appendix~\ref{D}.}

The integrand in \eqref{bla} decreases as a Gaussian with the distance to the
beamline. The Gaussian factor is the same as the one that
appears as a global factor in the energy-momentum tensor of the source beam (see
\cite{gaussstrahl} or appendix~\ref{A}), which implies that significant contributions to
$\Delta_+$ for the infinitely extended test ray are only
accumulated in regions where the energy distribution of the source
beam 
does not vanish. In addition, \eqref{blafin} shows that there is
no effect outside of a finite beam { 
when a  cut-off of the energy-momentum distribution is considered.}

\begin{figure}[H]\center
\includegraphics[scale=0.6]{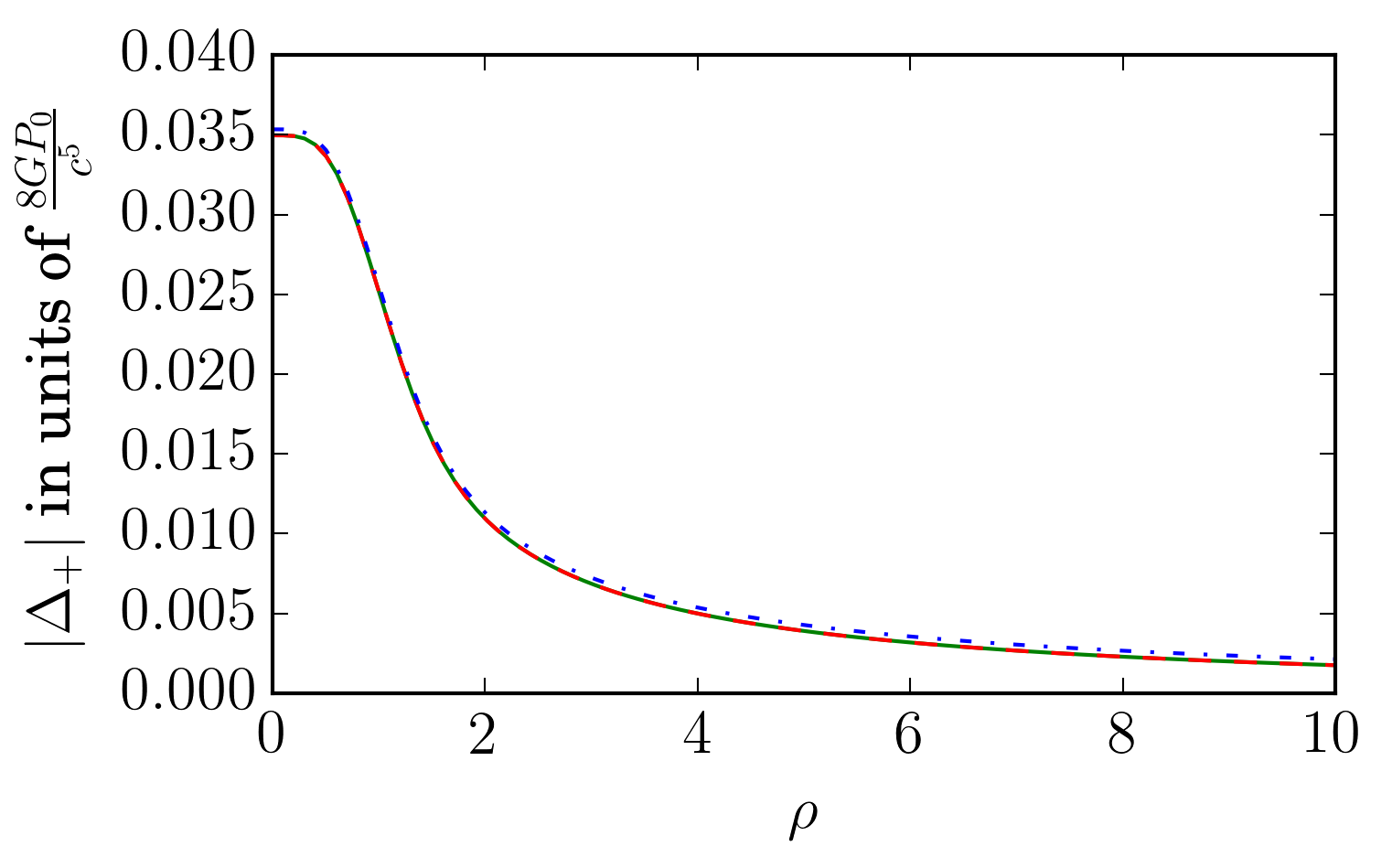}
\includegraphics[scale=0.6]{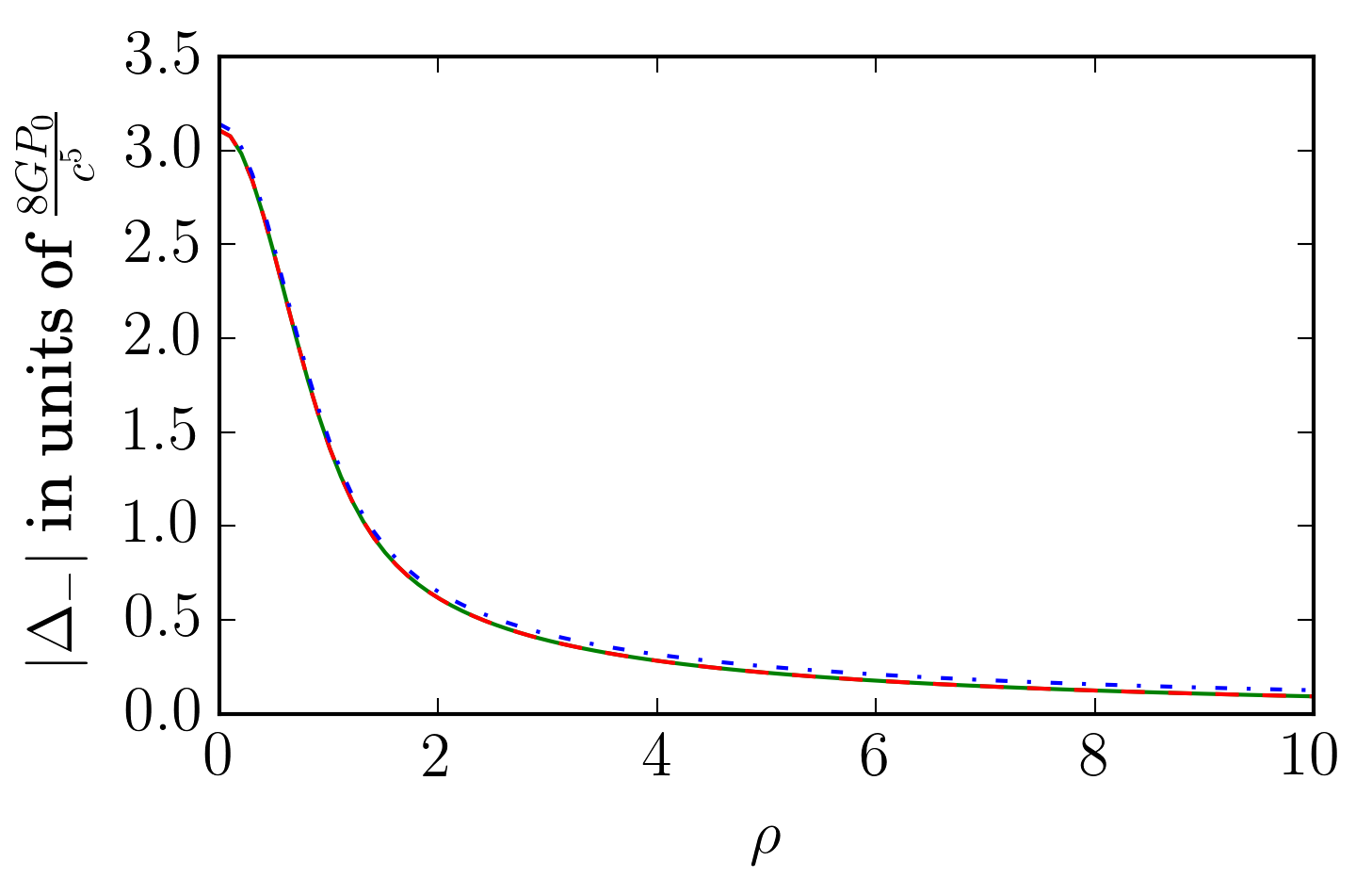}
\caption{\label{fg:deltacopara} The absolute value of the {
    polarization} rotation 
  angle $\Delta_{+}$ (upper graph) for  a
  parallel co-propagating light ray and $\Delta_{-}$ (lower graph) for
   a parallel counter-propagating { light ray}
  as a function of the transversal distance $\rho$ from the beamline. 
  The blue (dashed-dotted) line gives
  the rotation angle for the infinitely extended source beam and
  test ray. The green (unbroken) line gives the rotation angle for a source beam 
  with emitter and absorber at $\zeta =  -200$ and $\zeta = 200$, respectively,
  and infinitely extended test ray. The red (dashed) line gives the numerical values for 
  the same extensions of the test beam and a finitely extended test
  light-ray with  emitter (absorber) and absorber (emitter) at $\zeta =A=-600$ and $\zeta = B=
  600$, respectively, for the co-propagating (counter-propagating) beam. For the parameters given in the introduction,
  the factor $8GP_0/c^5$ is of the order $10^{-37}$. 
The plots show good   agreement between our results for finitely and
infinitely extended beams close to the beamline.   
}
\end{figure}

The sign of the rotation angle in equation (\ref{bla}) depends on 
$\lambda$, which specifies the handedness of the light in the source
laser-beam. The dependence of the rotation angle $\Delta_+$ on the
distance to the beamline is illustrated in the upper graph of
figure~\ref{fg:deltacopara}.  

For the counter-propagating test ray, we obtain in leading order (the first
order in $\theta$)
\begin{align}\label{kuchen}
\Delta_{-}&= { -\lambda\frac{8 GP_0\theta}{c^5 }
\int_{\alpha}^\beta d\zeta\,} |\mu|^2 e^{-2|\mu|^2\rho^2} 
\;.
\end{align}
for the finitely extended source beam and the infinitely extended test
ray. Equation~\eqref{kuchen} is derived with the same limiting procedures
as \eqref{bla}. Its version with finite radial
cut-off of $T_{\mu\nu}$ is given in \eqref{kuchenfin}. 
The integrand in equation~(\ref{kuchen}) decreases in the same way as the
one in equation~(\ref{bla}) with the same Gaussian factor with the
distance to the beamline that can be found as a global factor in the
energy-momentum tensor of the laser beam. We find that 
there are no significant contributions to the rotation angle $\Delta_-$ outside of the energy distribution for an infinitely
extended test ray (see equation (\ref{kuchenfin})) 
{ when introducing a  cut-off of the energy-momentum distribution in transversal direction.}
The dependence of the
rotation angle $\Delta_{-}$ on the distance to the beamline is
illustrated in the lower graph in figure~\ref{fg:deltacopara}. 
The two orders of magnitude larger values for $\Delta_-$ compared to
those for $\Delta_+$ arise due to the factor $\theta^2/8$ present in the expression for $\Delta_+$ but not in the one for $\Delta_-$ {(compare equations~\eqref{bla} and \eqref{kuchen}).}

\subsubsection{Transversally propagating test rays\label{orth}}
\begin{figure}[H]\center
\includegraphics[scale=0.17]{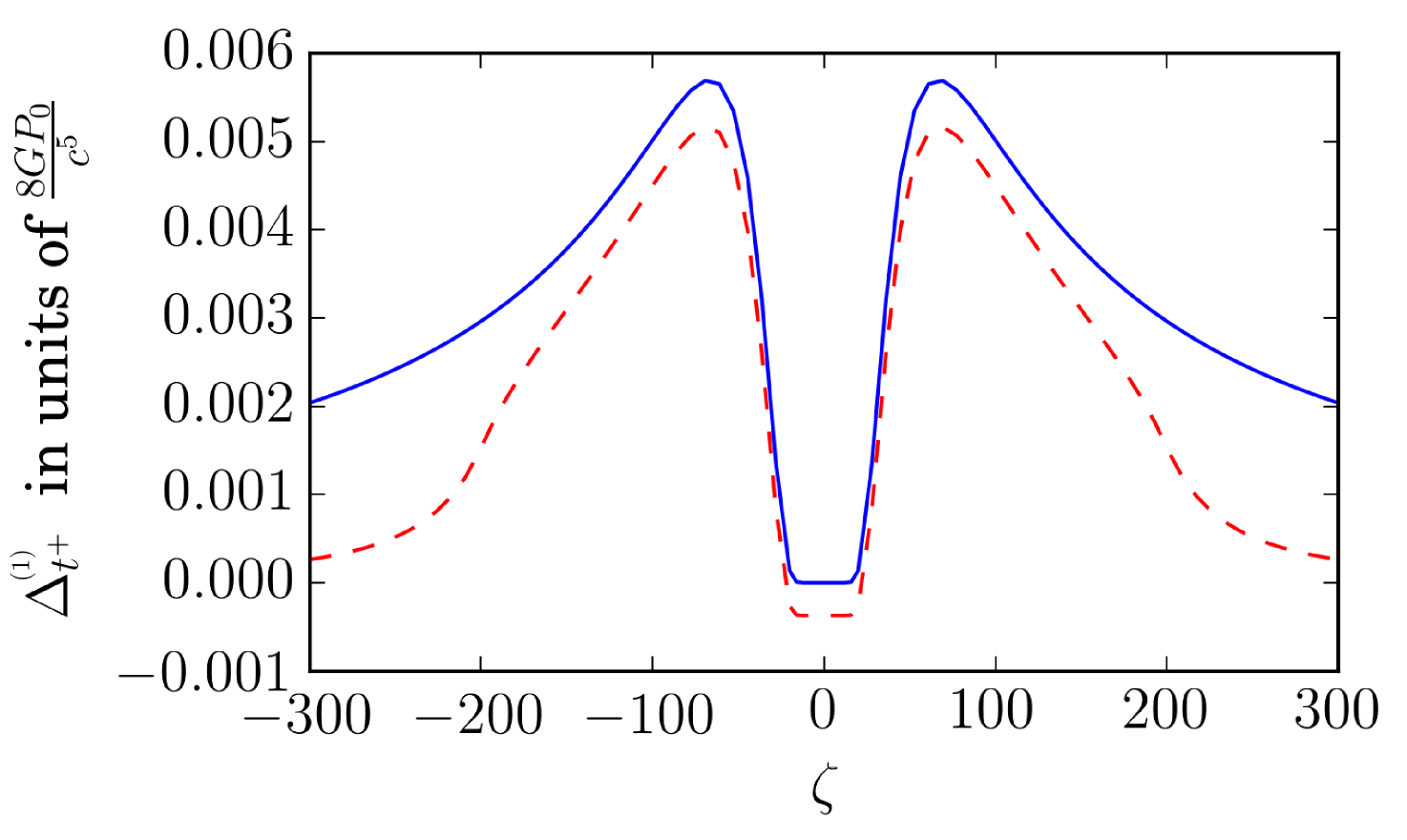}
\caption{\label{2firstorder}  First order contribution { (corresponding to the leading order effect of gravitational optical activity) } to the rotation
  angle $\Delta_{t^+}$ 
for the polarization vector of an transversally
propagating test ray with $\lambda=+1$: The blue, continuous line
  corresponds to the infinitely extended source beam, and the red,
  dashed line corresponds to the finitely extended source beam,
  emitted at $\mathcal{\alpha}=-200$ and absorbed at
  $\mathcal{\beta}=200$. 
  The test ray runs from $\xi = A=-600$ to $\xi=B=600$ at $\chi=10$.
  We find that the results for the
  infinitely extended source beam and test ray can be used to describe the effect in the
  case of the finitely extended source beam and test ray to some approximation for
  $\zeta$-positions that are in between emitter and absorber, but far
  from them. It can be seen that the rotation decreases fast at the ends
  of the finitely extended source beam.  
}
\end{figure}

The transversally propagating test ray is described by the
initial tangent $\dot\gamma_\pm=(1,\pm (1-f^\pm),0,0)$. Due to the
same argument as before, we do not have to take into account the
parameter $f^\pm$. For the rotation angle of the polarization vector,
we obtain for the infinitely extended source beam and infinitely
extended test ray (see appendix~\ref{D} for the detailed derivation)
including terms up to first order {
\begin{align}\A
\Delta_{\rm t^\pm}
=& \pm\frac{1}{2w_0^2}\int_{-\infty}^\infty d\xi \,\partial_\chi h_{\tau\zeta}^{(0)}
+\frac{\theta}{2w_0^2}\int_{-\infty}^\infty d\xi\; 
\partial_\chi h_{\xi\zeta}^{(1)}\\ 
=& \pm\frac{4\pi G P_0 }{c^5}\;{\rm erf}\left(\sqrt{2}|\mu|\chi\right)
+\lambda\frac{2\sqrt{2\pi}GP_0\theta}{c^5}|\mu|e^{-2|\mu|^2\chi^2} \;.\label{t}
\end{align}}
Let us denote the first term in equation~(\ref{t}) 
as $\Delta^{(0)}_{t^\pm}$ and
the second term as $\Delta^{(1)}_{t^\pm}$. Then, we find that 
$\Delta^{(1)}_{t^\pm}=\frac{\lambda\theta}{4} \partial_{\chi} \Delta^{(0)}_{t^+}$. 
 One might think that the symmetry of the beam
geometry 
implies that $\Delta^{(0)}_{t^\pm}$ should vanish at $\zeta =0$ 
as the term is independent of the helicity of the source beam.
However, the symmetry is broken due to the direction of
propagation of the source laser-beam. This can also be seen from
the fact that only the $\tau\zeta$-component of the metric perturbation
contributes to the effect, which would vanish for a massive
medium at rest (see for example the Levi-Civita metric for
an infinitely extended rod of matter \cite{levicivita_1919}).
The effect is similar to the
deflection of a transversally propagating test ray, which is both deflected
radially towards the laser beam as well as in $\zeta$-direction
\cite{tolman}. To illustrate the $\zeta$-dependence of $\Delta^{(1)}_{t^\pm}$, a
numerical evaluation and a comparison to results for a finitely extended source beam (see the following subsection) are given in
figure \ref{2firstorder}.

\begin{figure}[H]\center
\includegraphics[scale=0.55]{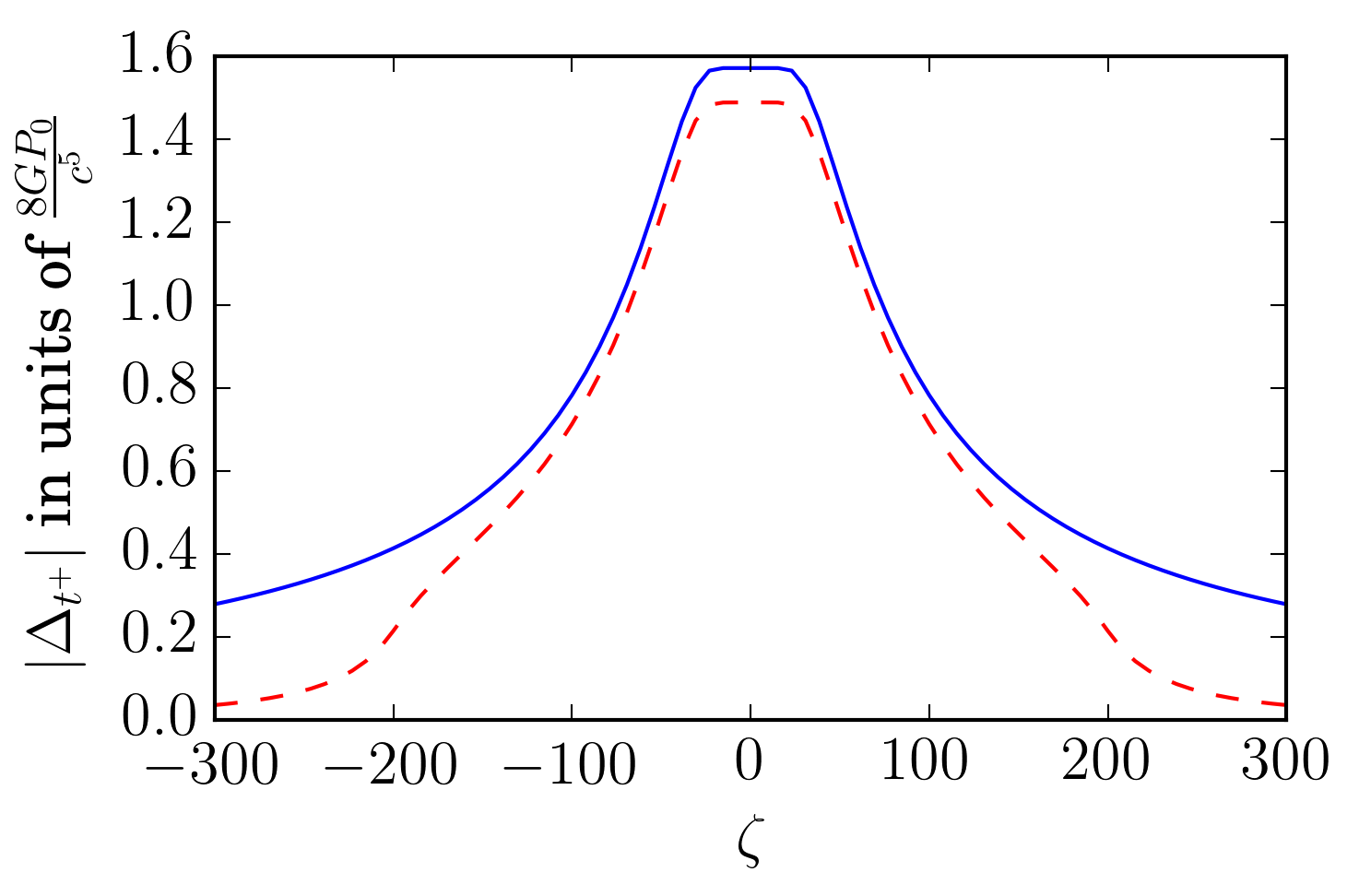}
\includegraphics[scale=0.55]{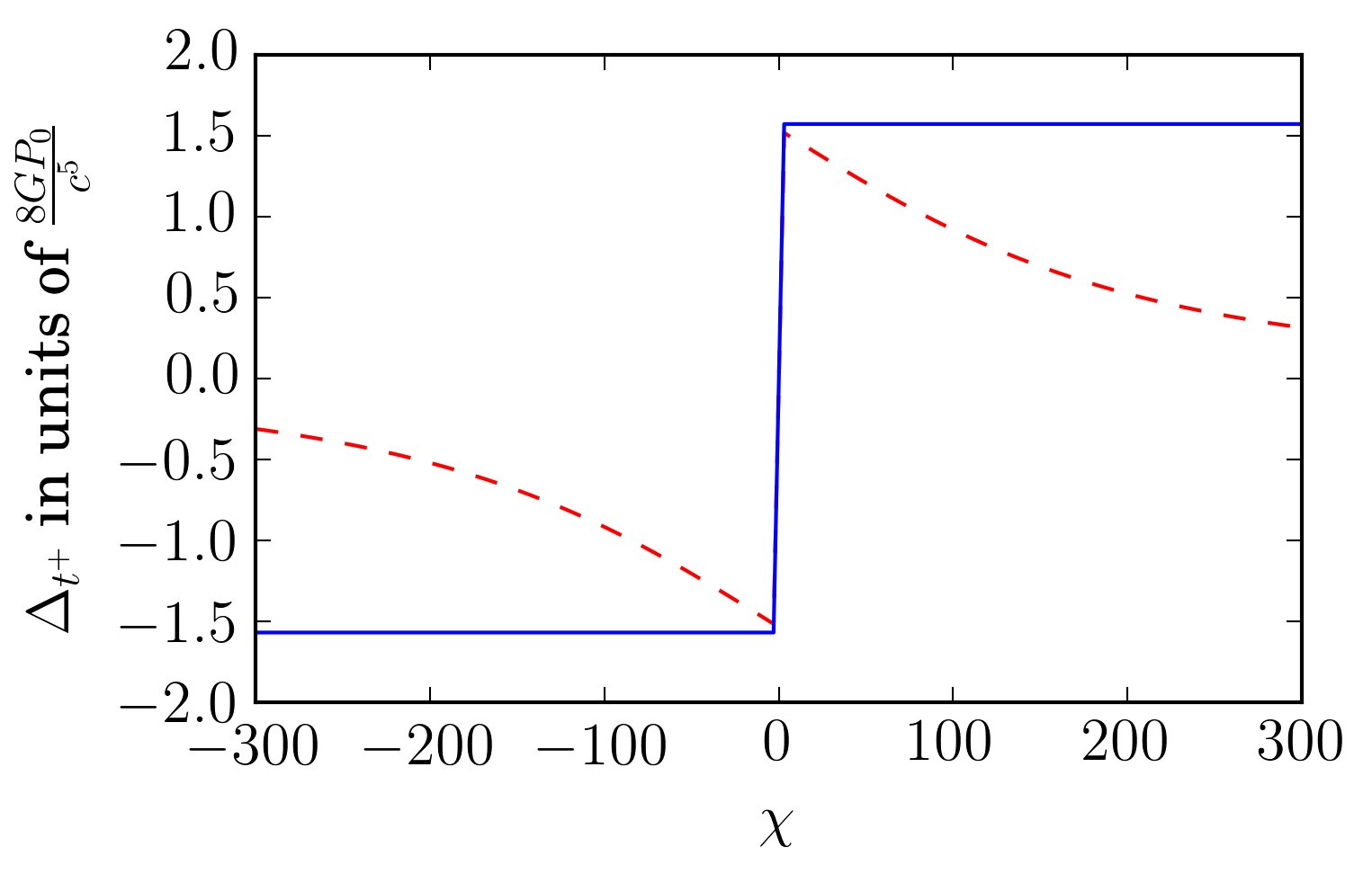}
\caption{\label{2} The rotation angle $\Delta_{t^+}$ (zeroth and first
  order) for the 
  polarization vector of an transversally propagating test ray:
  The blue, plain line corresponds to the infinitely extended source
  beam, and the red, dashed line corresponds to the finitely extended
  source beam, emitted at $\mathcal{\alpha}=-200$ and absorbed
  at $\mathcal{\beta}=200$. For the finitely extended source
  beam, the test ray is emitted at $\xi = A=-600$ and
  absorbed at $\xi=B=600$. In the first plot, $|\Delta_{t^+}|$ 
  is given as a function of the
  coordinate $\zeta$ along the beamline for $\chi=10$. For the parameters given in the
  introduction, the factor $8GP_0/c^5$ is of the order $10^{-37}$. We
  find that the results for the infinitely extended beam approximate
  those for the finitely extended beam for $\zeta$-positions in
  between emitter and absorber that are far from emitter and
  absorber. It can be seen that $|\Delta_{t^+}|$ decays quickly
  outside of the range of the finitely extended source beam and test ray in contrast to
  $|\Delta_{t^+}|$ for the infinitely extended ones
  that always overlap. 
  In both cases, the maximal
  effect is obtained close to $\zeta=0$.  
In the second plot, the angle $\Delta_{t^+}$ is given as a function of
$\chi$ at $\zeta=0$. For large values of $\chi$, it reaches 
a constant value for the infinitely extended source beam and test ray (undashed,
blue) and decreases for the finitely extended source beam  and test ray (dashed,
red). A dependence on $\chi$ as $1/\chi^2$ is found for larger values of
$\chi$ using a multipole expansion presented in appendix~\ref{mult}.
}
\end{figure}

The first and the second term in equation (\ref{t}) are fundamentally
different in their dependence on the variable $\chi$, which can be
interpreted as the impact parameter of the scattering of the test
light-ray with respect to the source beam. 
$\Delta^{(1)}_{t^\pm}$
is proportional to the same Gaussian function of $\chi$ that appears
as a global factor in the energy-momentum tensor of the source beam
for $\xi = 0$, which means that it vanishes if there is no overlap of
the source beam and the test ray in the same way as in the case
of $\Delta_+$ and $\Delta_-$ above. 
Instead, the first term in equation (\ref{t}) vanishes at $\chi=0$ and
saturates for large values of $\chi$ at a finite value, see figure
\ref{2} for plots showing numerical values for the first term in
(\ref{t}) and for the finitely extended source beam. 
Up to numerical factors of order 1, the prefactors in  equations
(\ref{bla}), (\ref{kuchen}), and \eqref{t} can be interpreted as the
ratio of the 
power $P_0$ of the source laser-beam to the Planck power
${E_p}/{t_p}={E_p^2}/{\hbar}$, where $E_p=\sqrt{{\hbar c^5}/{G}}$ is
the Planck energy, which explains the smallness of the effect.

\subsection{\label{finite}Finite vs. infinite source beams and test rays and the long range behavior}
For potential future experiments, finitely extended test-rays
are relevant.  It may even not be possible to realize
extensions of the test ray much larger than that of the source beam or
one may need to know details about the decay of the effect
for large distances from the beamline. 
It should then be kept in mind that \eqref{eq:rotpleb} holds for test
rays that
begin and end in flat spacetime. This is a condition which can   be 
fulfilled only approximatively for a finitely extended test-ray.
Furthermore, only under this condition has the rotation of the
polarization a 
clear physical, coordinate-invariant meaning. 
To give a 
physical meaning to the rotation angle for a finitely extended
test-ray, a physical reference system may 
be considered that extends from emitter to absorber. 
To this end, matter properties of the reference
system like its density and stiffness have to be taken into account
to obtain a reliable result. This is very similar
to the considerations we made in \cite{Raetzel:2018freq} for
the frequency shift of 
an optical resonator in a curved spacetime. We do not
follow such an approach in this article.

Here we rather focus on the question under which conditions
equation~\eqref{eq:rotpleb}, when integrated over a finitely extended test
ray, 
leads to results comparable to those of the infinitely extended
test-ray.  We will find that sufficiently close to the beamline the
results from the finite integration can be very close to those of an
infinite test-ray, which suggests that the latter, rigorous results with
clear physical meaning, also remain valid for experiments using a
finitely extended test-ray close to the source beam.  The situation is
quite different, however, in the far field, where results from the
finite source beam and the infinitely extended one, both evaluated
using \eqref{eq:rotpleb}, can differ siginificantly. 
This can be shown with a multipole expansion  
based on equation (\ref{eq:primeli}) or by numerically evaluating
equation (\ref{eq:primeli}). The basic 
expressions for the numerics are given in appendix~\ref{C} and the multipole expansion is performed in apendix~\ref{mult}.  Here we briefly
discuss both approaches and the main results.

The numerical values for the rotation angle for
  finitely extended 
   test rays and source beams presented in
figure~\ref{fg:deltacopara} are obtained from equations
  (\ref{58}) and (\ref{59}). 
The derivative in equation
(\ref{eq:rotpleb}) acting on the metric perturbation is shifted to the
energy-momentum tensor by pulling it into the integral, using the
symmetry of the function $|\vec{x}-\vec{x}'|$ to replace the
derivative for an un-primed coordinate by a derivative for a primed
coordinate and partial integration. 
The resulting
expressions are evaluated using Python and the scypy.integrate.quad
and scypy.integrate.romberg methods. The results for the
finitely extended beam and those for the infinitely extended beam are
very similar close to the beamline, see figure~\ref{fg:deltacopara}. 
The region in the $\xi$-$\chi$-plane
containing most of the energy of the source beam can be defined by a
drop of its intensity by a factor $e^{-2}$, which implies a radius
$w(\zeta)=\sqrt{1+(\theta\zeta)^2}$ of that region. In standard
notions $w(\zeta)$ is called the width of the beam as a realistic beam
is never infinitely extended in the transversal direction and is
usually considered to extend only on length scales of the order of
$w(\zeta)$. Equations (\ref{bla}) and (\ref{kuchen}) imply that there
is only a significant rotation angle accumulated along an infinitely
extended test 
ray if the latter overlaps with the region bounded by
the source beam's width, as the integrands in equations (\ref{bla})
and (\ref{kuchen}) are proportional to the same Gaussian function that
can be found as a global factor in the energy-momentum tensor of the
source beam. 
In the following, we will call this situation an overlap of the test
light ray and the source beam.  That $\Delta_-$ and $\Delta_+$ are only non-zero for an
overlap of test ray and source beam is confirmed by equations (\ref{blafin}) 
and (\ref{kuchenfin}), where a cut-off of the source beam's
energy-momentum distribution is considered. For $\theta\zeta \gg
1$,  we find that  $w(\zeta)\approx \theta\zeta$. 
Therefore, a test ray at $\rho \gg 1$
 overlaps with the source beam only in regions where
$|\theta\zeta| > \rho$. For the infinitely extended 
source beam and test ray, there  is always an overlap, 
but it does not need to be the case if at least one of
  the two beams has finite length. 

\begin{figure}[H]
\includegraphics[scale=0.9]{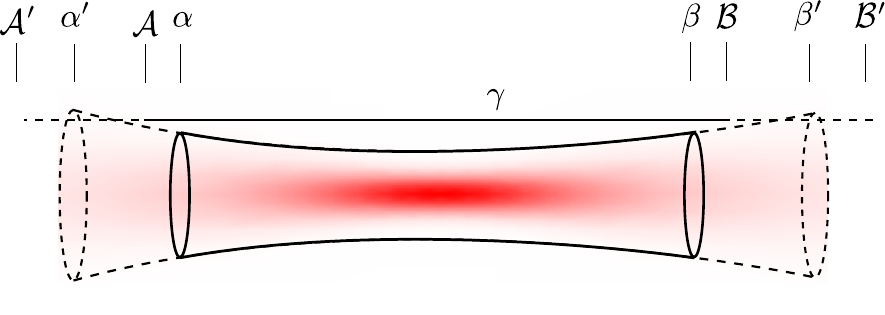}
\caption{Illustration of the overlap of the test ray with the
  source laser-beam: A test ray may overlap with the source
  laser-beam only if the latter is long enough. In the illustration,
  the path of the test ray is labelled by $\gamma$ and starts
  and ends at $\mathcal{A}$ and $\mathcal{B}$ respectively for the
  short source laser-beam (starting and ending at $\alpha$ and $\beta$
  respectively) or at $\mathcal{A}'$ and $\mathcal{B}'$ for the long
  source laser-beam (starting and ending at $\alpha'$ and $\beta'$
  respectively). \label{abf}} 
\end{figure}

\begin{figure}[H]\center
\includegraphics[scale=0.6]{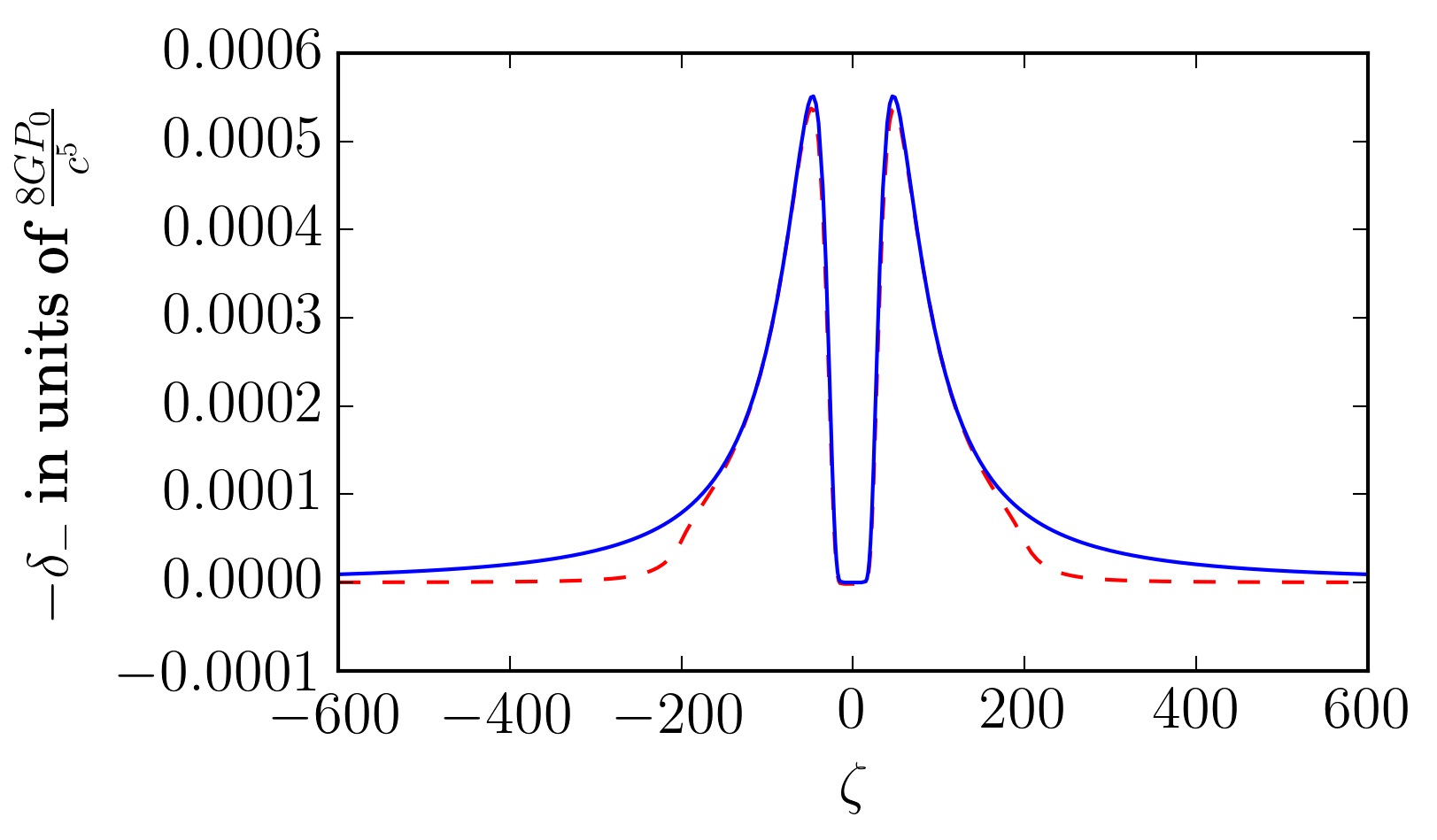}
\caption{ \label{fg:deltaminloc}
The function $-\delta_{-}$ (the
 integrand in equation (\ref{kuchen})) for the polarization vector of
 the parallel counter-propagating light ray is plotted as a
 function of the coordinate along the beamline  $\zeta$ for a distance from the beamline  $\rho =
 10$. The blue (unbroken) line gives the rotation angle for the
   infinitely extended source beam and test ray as in equation
 (\ref{kuchen}). The red (dashed) line gives the
 numerical values for a finitely extended source beam with emitter and
 absorber at $\alpha = -200$ and $\beta = 200$, respectively, based on
 (\ref{eq:primeli}). It can be seen
 that $\delta_{-}$ decays quickly outside of the range of the finitely
 extended beam in
 contrast to $\delta_{-}$ for the infinitely extended source beam,
 which continues to decay like $1/\zeta^2$ for large $\zeta$ just as
 the source beam's energy density. The
 region left of the steep descent around $\zeta \sim -70$ and
 the region right of the steep ascent around $\zeta \sim 70$
 correspond to the overlap regions of source beam and test
 light-ray. In the case of an infinitely extended source beam, these
 regions are infinitely extended. In the case of a finitely extended
 source beam, the overlap regions end at the end of the source beam as
 can be seen with the steep ascent close to $\zeta = -200$ and the
 steep descent close to $\zeta = 200$ for the red (dashed) curve.  
}
\end{figure}

Note that for large values of $\theta\zeta$, the energy-density of the
source laser-beam is proportional to $(\theta\zeta)^{-2}$ (while
transversally it decreases as a Gaussian with the distance
to the beamline). The same is true for the integrands in
equations~(\ref{bla}) and (\ref{kuchen}). Therefore, 
$\Delta_\pm$ in 
equations~(\ref{bla}) and (\ref{kuchen}) are approximately
proportional to $1/(\theta\zeta)$ evaluated at the boundaries of the
regions where test ray and source beam overlap. For the
infinitely extended beams, this implies that the rotation angles in
equations~(\ref{bla}) and (\ref{kuchen}) are approximately
proportional to $1/\rho$ for large $\rho$. 
The proportionality of $\Delta_-$ and $\Delta_+$ to $1/\rho$ holds as well for finitely
extended source 
beams if $\rho \ll -\theta \alpha$ or $\rho \ll \theta
\beta$. For larger values of $\rho$, there is no overlap
of test ray and source beam (this is illustrated in figure~\ref{abf}
and figure~\ref{fg:deltaminloc}). Then,
$\Delta_-$ and $\Delta_+$ decay proportional to
$e^{-\Sigma\rho^2}/\rho^2$ and $e^{-\Sigma\rho^2}$, respectively,
where $\Sigma=2/(\theta\alpha)^2$ for $\alpha \ge -\beta$ or
$\Sigma=2/(\theta\beta)^2$ for $\beta \ge -\alpha$, as shown in
equation (\ref{eq:decay}) and equation (\ref{eq:decaypl}),
respectively. 
The behavior for large distances from the beamline and
finitely extended test rays can be analysed with a
multipole expansion, assuming that the source term in the form of the
derivatives of the energy-stress tensor can be effectively cut-off at
$w(\zeta)$.  This is presented in appendix~\ref{mult}. One finds 
that for $\Delta_\pm$ the lowest contributing moment is a
quadrupole leading to a $1/\rho^3$ decay for finite $B=-A$.  At the
same time, the prefactor of these terms decay as $1/B^2$ for $B\gg
\rho$.  Higher multipoles lead to an even faster decay, both with $\rho$
and $B$.  Hence, in the case of a finitely extended source beam and an
infinitely extended test ray that does not overlap with the source
beam, one expects to recover the fast decay of $\Delta_\pm$ with $\rho$
obtained in equations~(\ref{bla}) and (\ref{kuchen}).
However, a resummation of the multipole expansion would be needed to
find out its functional form.  This is beyond the scope of the present
investigation.  Nevertheless, the analysis makes clear that
$\Delta_\pm$ sensed by a finitely extended test ray 
in the far-field regime is not
captured accurately by the results from the idealized infinitely
extended test ray for the cases considered.

%======================================================================================
For the transversal test ray, 
the $\chi$-dependence 
 of $\Delta_{t^\pm}$ for $\chi \gg 1$ 
changes drastically 
 for the finitely extended source beam 
compared to the infinitely  extended one. In particular, the result 
 that the first term in equation~(\ref{t}) does not vanish for
 large distances 
from the beamline 
turns out to be
an effect of the infinite extension of source
beam and test ray. 
Alternatively, this can also be seen as
follows: 
As $\Delta^{(0)}_{t^\pm}$ is of zeroth order, it
remains present when describing the laser beam in the paraxial
approximation, in which the gravitational field of an infinitely extended
source beam 
has the form $h_{00} = h_{33}
= -h_{30} \propto \ln(\rho)$ (see \cite{aichelburg1971,raetzel_2016}
and consider an infinite pulse length or see \cite{bonnor_1969},
consider an energy distribution localized to the beamline, and subtract
the Minkowski metric from the resulting spacetime metric). From
equation (\ref{eq:rotpleb}) and for a transversal
infinitely extended
test ray, we immediately obtain a rotation angle proportional to
the first term in equation (\ref{t}). 
On  the other hand, 
the solution 
for the gravitational field 
for a finitely
extended source beam can be found in \cite{tolman}. { In
  appendix~\ref{G},} using this 
solution and an infinitely extended test ray, 
we obtain the
radial dependence of the rotation angle as $1/\chi$, and for a
finitely extended test ray, we find that the rotation angle is
proportional to $1/\chi^2$ for large $\chi$. 
This is corroborated by the multipole expansion, where we find a
monopole contribution responsible for the $1/\chi^2$ behavior to
zeroth order in $\theta$ for $\chi\gg B$. 
As function of $B=-A$ it saturates for large $B$ (i.e.~$B\gg\chi$) and
gives a $\beta/\chi$ behavior, see appendix~\ref{mult}.

Since $\Delta^{(1)}_{t^\pm}=\frac{\lambda\theta}{4} \partial_{\chi} \Delta^{(0)}_{t^+}$, 
we find that $\Delta^{(1)}_{t^\pm}$ decays as $1/\chi^3$ for finitely extended source beams
and test rays and as $1/\chi^2$ for finitely extended source beams and infinitely extended test rays. The corresponding multipole expansion is given in appendix~\ref{mult}.

%===============================================================
\subsection{Rotation of polarization and gravitational spin-spin coupling\label{spin}}

The rotation angles $\Delta_\pm$ as well as 
the first order contribution to
$\Delta_{t^\pm}$ are proportional to the helicity $\lambda$ of the
source laser-beam. As explained in the end of section~\ref{rot}, the
rotation angle is equivalent to a phase for circularly polarized test
light rays, which is given by $-\lambda_{\rm test} \Delta$. 
This phase contains the product of the helicities of the source
laser-beam and the test ray, $\lambda\lambda_{\rm test}$. 
Therefore, the
phase depends on the relative helicity of the two beams. This is
gravitational spin-spin coupling. 

We can consider the source beam as its own test beam,
$\lambda_\text{test}=\lambda$, such that $\lambda_\text{test}\Delta_+ = C_+$
where $C_+>0$ is a function that increases monotonously with
the end of the source beam at $\zeta=\beta$ (see \eqref{bla}). 
Since $C_+$ enters as a phase ${\rm Exp}(iC_+)$, 
it can be combined with the global plane wave factor 
at the end of the beam $\zeta=\beta$ as ${\rm Exp}(i\Phi)$ 
where $\Phi = 2(\beta - \tau)/\theta + C_+$. This leads to the locally modified wave number
$\tilde k = \partial_\beta\Phi =\left(2 + \theta\, \partial_\beta C_+\right)/\theta$ 
at $\zeta =\beta$.
Effectively, this leads to the interpretation of a locally modified dispersion
relation and an effectively reduced speed of light.
 This self-interaction effect
is proportional to the intensity of the electromagnetic field.
It is reminiscent of the apparent modification of the speed of light
found in \cite{braun_intrinsic_2017} based on the eikonal
approximation of the solution of the relativistic wave equation of a
light-beam in its own gravitational field.

%======================================================================================
\section{Faraday effect and optical activity\label{Ff}}

The electromagnetic Faraday effect is a non-reciprocal
phenomenon. Non-reciprocity means that the effect does not cancel when
the test ray propagates back and forth along the same path. 
We investigate this feature for its gravitational analogue.

The rotation angle given in equation (\ref{eq:rotpleb}) is defined with respect
to the propagation direction. Therefore, the absolute rotation accumulated
on the way back and forth through spacetime seen by an external 
reference system at the starting point of the test ray's trajectory
at spatial infinity is given by the difference between the
rotation angle acquired on the outbound trip and the one
acquired on the way back. For a tangent vector $t_0^\mu$ with $t_0^0=1$ and $t_0^a=d\,s^a$ with $d=+1$ for outbound and $d=-1$ for back propagation, we obtain from equation (\ref{eq:rotpleb}) the rotation angle
\begin{equation}\label{eq:rotplebsep}
 \Delta_{s,d}
= \frac{1}{2w_0^2} \int_{-\infty}^\infty d \tau\;
  s^a \epsilon_{abc} \partial_b (h_{cf}s^f + d h_{c\tau}) \;,
\end{equation} 
and therefore, the Faraday rotation for one roundtrip becomes
\begin{equation}\label{eq:faradaygen}
 \Delta_s^\mathrm{F}
= \Delta_{s,+} - \Delta_{s,-} = \frac{1}{w_0^2} \int_{-\infty}^\infty d\tau\;
  s^a\epsilon_{abc} \partial_b h_{c\tau} \;.
\end{equation}
We find that the gravitational Faraday effect is given by the spacetime-mixing 
component of the metric perturbation $h_{c\tau}$. In contrast, the first term in (\ref{eq:rotplebsep})
containing a purely spatial component of the metric perturbation does not 
depend on the propagation direction and cancels
on the way back and forth. This is the gravitational optical activity 
for a single trip
\begin{equation}\label{eq:opgen}
\Delta_s^\mathrm{Op} = \frac{\Delta_{s,+} + \Delta_{s,-}}{2}
= \frac{1}{2w_0^2} \int_{-\infty}^\infty d\tau\;
  s^a s^d\epsilon_{abc} \partial_b h_{cd}  \;.
\end{equation}

For the rotation due to the gravitational Faraday effect { after one
  roundtrip for the parallel} test ray, we obtain from equation
(\ref{pm}) { to leading order}  
\begin{align}\label{pap} \A
\Delta^\mathrm{F}_{+-} &= \Delta_{+} - \Delta_{-}\\
&= { -\frac{\theta}{w_0^2}\int_{-\infty}^\infty d\zeta}
 \Big(\partial_\chi { h^{(1)}_{\tau\xi} -\partial_\xi h^{(1)}_{\tau\chi}}\Big)\;.
\end{align}

Adding the rotations due to the transversal back and forth propagation leads to (the explicit expression is identical to twice the positive contribution of the first term in equation (\ref{t})),
\begin{align}\label{out}
\Delta^\mathrm{F}_{\rm t^+t^-} = \Delta_{\rm t^+} - \Delta_{\rm t^-}
&= \frac{1}{w_0^2}\int_{-\infty}^\infty  d\xi\;\partial_\chi h_{\tau\zeta}^{(0)}\;,
\end{align}
which means that the effect is of zeroth order. 
The contribution of gravitational 
optical activity is
given as { (to leading order { and for one direction of propagation})}
\begin{align} \A
\Delta^\mathrm{Op}_{+-} &= { \frac{\Delta_{+} + \Delta_{-}}{2}}\\
&= { -\frac{\theta}{2w_0^2}\int_{-\infty}^\infty d\zeta
 \Big(\partial_\chi h^{(1)}_{\zeta\xi} -\partial_\xi h^{(1)}_{\zeta\chi}\Big) }\;
\end{align}
for the parallel test rays, and
\begin{align}\label{out}
\Delta^\mathrm{Op}_{\rm t^+t^-} = { \frac{\Delta_{\rm t^+} + \Delta_{\rm t^-}}{2}}
&= { \frac{\theta}{2w_0^2}}\int_{-\infty}^\infty  d\xi\;\partial_\chi h_{\xi\zeta}^{(1)}
\end{align}
for the transversal test rays.  

 From the vanishing of $\Delta_+$ in first order in the metric perturbation, 
we deduce that the first order contributions of optical activity and the Faraday effect
to the polarization rotation accumulated along a parallel co-propagating test ray
have the same absolute value and cancel each other.
In contrast, the two contributions add for the counter-propagating test ray.
This situation can be compared to the result of Tolman et
al.~\cite{tolman}, which states that a test ray is not deflected in the
gravitational field of a source light-beam if it is parallel
co-propagating, while it is deflected if it is parallel
counter-propagating. It is the motion of the source of gravity that
breaks the symmetry; its motion with the speed of light leads to the
extreme case of equal absolute values of the two effects.

%===========================================================
\subsection{Spacetime-medium analogy\label{H} }

 The above identification of the two distinct rotation effects and the
different types of components of the metric perturbation can be compared with the formal analogy
of electrodynamics in linear dielectric media and electrodynamics in a weakly 
curved spacetime employed in \cite{plebanski_1960}. In particular, 
Maxwell's equations in a curved spacetime can be rewritten such that they have their usual form in dielectric media \cite{plebanski_1960}
\begin{eqnarray}
	-\partial_t D_a + \epsilon_{abc} \partial_b H_c = i_a \,, \\
	\partial_t B_a  +  \epsilon_{abc} \partial_b E_c = 0 \,,\\
	\partial_a D_a = \rho \,,\quad \partial_a B_a = 0
\end{eqnarray} 
by identifying the components of the field strength tensor as
\begin{eqnarray}
	E_a &=& c F_{a0}\\
	B_a &=& \frac{1}{2} \epsilon_{abc} F_{bc}\\
	D_a &=& \varepsilon_0 c \sqrt{-g} g^{0\mu}g^{a\nu} F_{\mu\nu}\\
	H_a &=& \frac{1}{2\mu_0} \epsilon_{abc} \sqrt{-g} g^{b\mu}g^{c\nu} F_{\mu\nu}\,
\end{eqnarray} 
and $i_a = \sqrt{-g} j^a$ and $\rho = \sqrt{-g} j^0/c$, where $j^\mu$ is the
current density, $\sqrt{-g}$ is short for $\sqrt{-\mathrm{det}(g)}$ and $g$ is the spacetime metric. This leads to an effective constitutive law
\begin{eqnarray}\label{eq:constitell}
	D_a &=& \varepsilon_{ab} E_b  + \epsilon_{abc} \Xi_b H_c\,,\\
	B_a &=& \mu_{ab} H_b + \epsilon_{abc} \Upsilon_{b} E_c\,,
\end{eqnarray}
where \footnote{Note that what we define as $\epsilon$ is the effective permittivity tensor in contrast
to \cite{plebanski_1960}, where the identity is subtracted.}
\begin{eqnarray}
	\varepsilon_{\rm{r},ab} &=& \mu_{\rm{r},ab} =-\frac{\sqrt{-g}}{g_{\tau\tau}} g^{ab}\,,\\
	  \Xi_{a}&=&-\Upsilon_{a}= \frac{g_{\tau a}}{c g_{\tau\tau}}  \,,
\end{eqnarray}
where $\varepsilon_{\rm{r},ab} = \varepsilon_{ab}/\varepsilon_0$ and $\mu_{\rm{r},ab}=\mu_{ab}/\mu_0$ are the relative permittivity tensor and relative permeability tensor, respectively.
In the above identification, purely spatial components of
the metric perturbation lead to non-trivial effective relative permittivity and permeability tensors while space-time mixing components of the metric perturbation lead to effective non-vanishing magneto-electrical mixing terms. Note the equivalence of the effective relative permittivity and permeability tensors, denoted as impedance matching \cite{Leonhardt_2006}, which is usually not encountered in materials.

In linearized gravity using $\eta=w_0^2\mathrm{diag}(-1,1,1,1)$ and the set of coordinates $\tau,\xi,\chi$ and $\zeta$, we find
\begin{eqnarray}
	 \varepsilon_{\rm{r},ab} &=& \mu_{\rm{r},ab} =\delta^{ab}\left(1+\frac{h_{\tau\tau} + \delta^{cd}h_{cd}}{2w_0^2}\right) - \frac{h_{ab}}{w_0^2}\,,\\
	  \Xi_{a}&=&-\Upsilon_{a}= -\frac{h_{\tau a}}{c w_0^2}  \,.
\end{eqnarray}
We obtain for the gravitational
Faraday rotation angle from equation (\ref{eq:faradaygen})
\begin{eqnarray}
\Delta_s^\mathrm{F}
\A  &=& - c\int_{-\infty}^\infty d\tau\; s^m  \epsilon_{mab}\partial_a\Xi_{b}\\
&=& - c\int_{-\infty}^\infty d\tau\; s^m  (\nabla\times \Xi)_{m} \;.
\end{eqnarray}
We see that the gravitational Faraday rotation is induced by the curl of the magneto-electrical mixing vector $\Xi$; it is a magneto-optical effect like its analogue in dielectric media. For the gravitational optical activity, we obtain from equation (\ref{eq:opgen})
\begin{eqnarray}
\Delta_s^{\rm{Op}}
\A &=& -\frac{1}{2} \int_{-\infty}^\infty d \tau\, s^m s^d \epsilon_{mbc} \partial_b \varepsilon_{\rm{r},cd}  \\
&=& -\frac{1}{2} \int_{-\infty}^\infty d \tau\, s^a s^b \left(\nabla\times\varepsilon_\mathrm{r}\right)_{ab} \;.
\end{eqnarray} 
We find that the gravitational optical activity is induced by a
non-vanishing curl of the 
lines (or columns) of the
permittivity tensor (and, equivalently, the permeability tensor). 
In analogy to the optical activity in dielectric media, the gravitational optical activity does not mix electric and magnetic fields.

For the gravitational field of the laser beam in appendix~\ref{A}, we obtain evaluated up to first order in $\theta$
\begin{eqnarray}
	 \varepsilon_{\rm{r},ab}&=& \mu_{\rm{r},ab} = \left(
	\begin{array}{ccc}
		1 + \frac{h^{(0)}_{\zeta\zeta}}{w_0^2} & 0 & -\frac{\theta h^{(1)}_{\xi\zeta}}{w_0^2} \\
		0 & 1 + \frac{h^{(0)}_{\zeta\zeta}}{w_0^2}  & -\frac{\theta h^{(1)}_{\chi\zeta}}{w_0^2} \\
		-\frac{\theta h^{(1)}_{\xi\zeta}}{w_0^2} & -\frac{\theta h^{(1)}_{\chi\zeta}}{w_0^2} & 1 
	\end{array}\right)\\
	\label{eq:magelexpl} \Xi_{j}&=&-\Upsilon_{j} 
	= -\frac{\theta h^{(1)}_{\tau j}}{c w_0^2}\,,\, \rm{where}\,\,\, j \in \{\xi,\chi\} \\
	\Xi_{\zeta }&=& -\Upsilon_{\zeta} = - \frac{ h^{(0)}_{\tau\zeta}}{c w_0^2}\,,
\end{eqnarray}
and we recover the results in equations (\ref{pap})-(\ref{out}).

%======================================================================================
\section{Test rays in cavities}\label{cavity}

In a one-dimensional cavity containing light that propagates back
and forth, the effect associated with gravitational optical
activity cancels while the gravitational Faraday effect adds up. 
{ In a ring cavity or an optical fiber coiled around the
{ beamline,} the full polarization rotation is accumulated
and the gravitational Faraday effect represents the leading
order effect. For the case of a transversally oriented
ring cavity, a situation can be created in which the 
Faraday effect vanishes and only the gravitational optical
activity accumulates. }

%==================================================================================
\subsection{Parallel linear cavity}

\begin{figure}[H]\center
\includegraphics[scale=0.6]{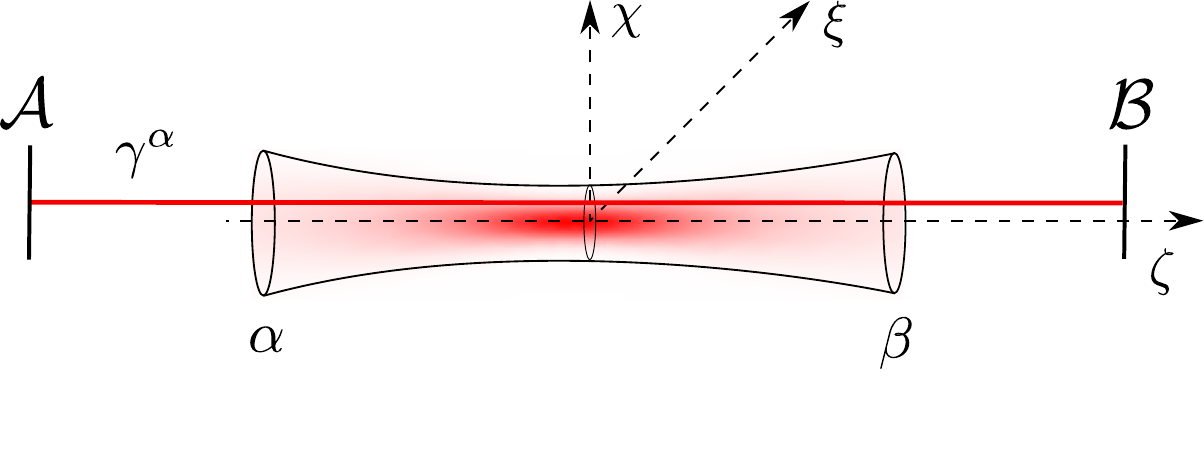}
\caption{\label{c}Schematic illustration of the parallel cavity in the
  gravitational field of the laser beam: The source laser-beam starts
  at $\alpha$ and ends at $\beta$. The test ray propagates on
  the worldline $\gamma$ between the mirrors $\mathcal{A}$ and $\mathcal{B}$ of the cavity. The  Faraday
  effect adds up after each roundtrip, while the rotation associated
  with gravitational optical activity vanishes. } 
\end{figure}

We consider a cavity consisting of two mirrors between which the light
propagates back and forth, with the axis of the cavity oriented
parallel to the beamline and at a distance $\rho$ from the beamline.   
The setup is illustrated in figure~\ref{c}.
Up to third order in $\theta$, the light travels undeflected from
$\zeta=A$ to $\zeta=B$ and picks up a small
deflection of zeroth order in $\theta$ when travelling from
$\zeta=B$ to $\zeta=A$. 
The deflection vanishes when the light ray propagates at the center of
the source beam, at $\rho=0$. In this case only the angle due to the
Faraday effect accumulates. For one back and forth propagation, it is
given by equation (\ref{pap}). 
Letting the light propagate during the time $\tau=LF/(\pi c)$, where
$F$ is the finesse of the cavity, the total angle of rotation is given
by $\Delta_{\rm  cav, +-}= \Delta_{+-}^\text{F}F/(2\pi)$.  
For a cavity of finesse $F=10^6$ {\cite{dellavalle_2014}} and the
parameters given in the introduction, the rotation angle is of the
order of magnitude $\Delta_{\rm cav, +-}\sim \pm 10^{-32}\,{\rm
  rad}$. For a cavity at distance $\rho>0$ from the center of the
laser beam, the effect is smaller, and one has to take into
consideration the deflection when the test ray is
counter-propagating to the source laser-beam. 

%==================================================================================
\subsection{Transversal linear cavity}

\begin{figure}[H]\center
\includegraphics[scale=0.7]{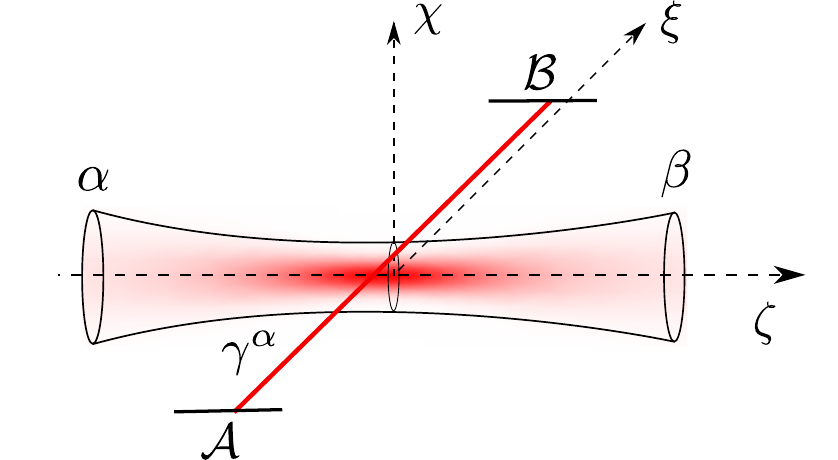}
\caption{\label{tr}Schematic illustration of the transversal cavity in the gravitational field of the laser beam: The test ray propagates along the worldine $\gamma$, marked as a red line, and is reflected at the mirrors $\mathcal{A}$ and $\mathcal{B}$. The source laser-beam is emitted at $\zeta=\alpha$ and absorbed at $\zeta=\beta$. The Faraday effect adds up after each roundtrip, while the rotation associated with gravitational optical activity vanishes. }
\end{figure}

Rotating the parallel cavity by ninety degrees, we obtain a
transversal cavity, as illustrated in figure~\ref{tr}.
Analogously to the parallel cavity, one finds that the total angle of rotation is given by $\Delta_{\rm cav,t^+t^-}=\Delta_{\rm t^+t^-}^\text{F}F/(2\pi)$. For a finesse of $F\sim 10^6$ and the parameters given in the introduction, it is of the order $\pm { 10^{-32}}\,{\rm rad}$. 

%%==================================================================================
\subsection{Ring cavity}

\begin{figure}[H]\center
\includegraphics[scale=0.7]{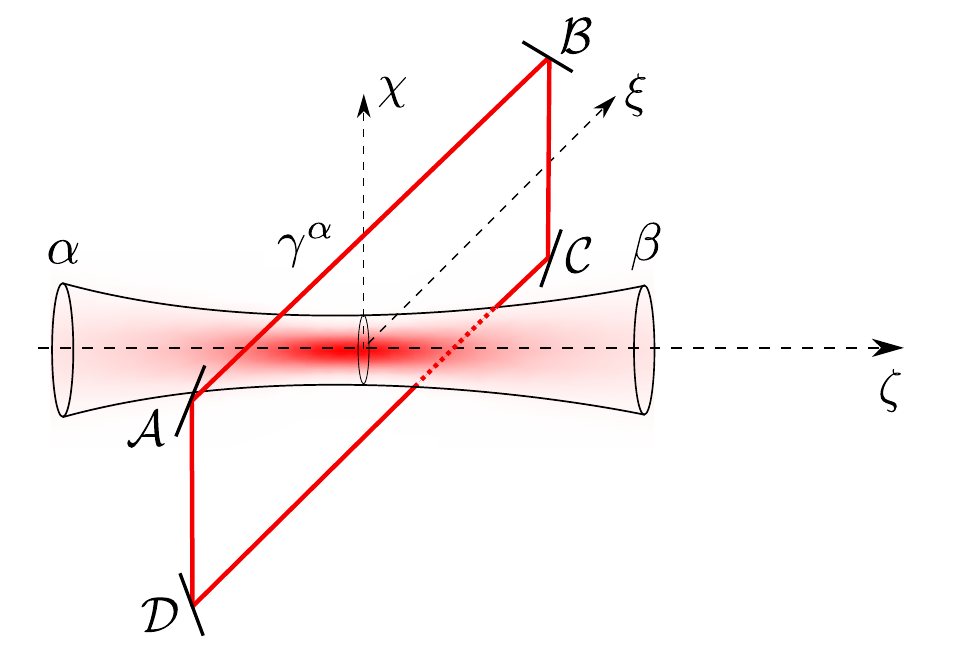}
\caption{\label{fig}
  Schematic illustration of the ring cavity setup: The
  test ray propagates along the path $\gamma$ and is reflected
  at the mirrors $\mathcal{A}$, $\mathcal{B}$, $\mathcal{C}$ and
  $\mathcal{D}$. The source laser-beam is emitted at $\zeta=\alpha$ and
  absorbed at $\zeta=\beta$. A similar situation can be created
  with a test ray in a wave guide that is wound many times
  around the source beam.}
\end{figure}

In order to measure the polarization rotation including the contribution due to optical activity for the transversal light ray, we consider a ring cavity: The light propagates
from { $\mathcal{A}$ at $(\xi,\chi,\zeta)=(-\infty,\chi_1,0)$, to
$\mathcal{B}$ at $(\xi,\chi,\zeta)=(\infty,\chi_1,0)$, to $\mathcal{C}$
at $(\xi,\chi,\zeta)=(\infty,\chi_2,0)${, where $\chi_1$ and $\chi_2$
  have opposite sign,} 
to $\mathcal{D}$ at $(\xi,\chi,\zeta)=(-\infty,\chi_2,0)$} and back to $\mathcal{A}$. The
$\pm\infty$ can be replaced by distances from the beamline much larger
than $\beta$. The polarization rotation accumulated when propagating from $\mathcal{A}$ to
$\mathcal{B}$ and from $\mathcal{C}$ to $\mathcal{D}$ add up. The setup is illustrated
in figure~\ref{fig}. 
The rotation of polarization after one roundtrip is given by twice the 
expression in equation (\ref{t}) for { $\chi_1 \sim 1$ and $\chi_2 \sim -1$.} For $\chi_1 \gg \beta$, $\chi_2 \gg -\beta$ 
and $\alpha = -\beta$, we have shown that the effect decays as $\beta/\chi^2$ in appendix~\ref{mult}. 
As the first term in equation (\ref{t})  { corresponding to the gravitational
Faraday effect} is of zeroth order in $\theta$, 
it does not depend on the beam waist for the fixed wavelength given by $\pi\theta w_0$.
This means that the beam has to be long, but it does not need
to be focused. Again for a finesse of $F=10^6$ and the parameters given in the
introduction, the rotation is of the order of magnitude $\Delta_{\rm
  t^+}F/(2\pi)\sim 10^{-32}\,{\rm rad}$.  

For $\chi_1=0$ and $\chi_2=-\infty$ or at least $-\chi_{ 2}$ very large,
we find that the polarization rotation due to the Faraday effect vanishes
(see also equation (\ref{eq:Deltat0}))
and the rotation due to gravitational optical activity remains
(see also equation~\eqref{hib}). Then, the accumulated effect is
by one order smaller than that 
due to the Faraday effect at $\chi_1=\chi_2>1$.

A ring cavity can also be used to amplify the rotation angle of the
polarization the parallel co-propagating test ray acquires: Since it
is not deflected, one can let the light ray pass through the
gravitational field $N$ times just in the direction of propagation of
the source beam, such that the effect is amplified by a factor $N$.

%==================================================================
\subsection{Measurement precision of the rotation angle\label{pres}}

The rotation angle $\Delta$ is experimentally inferred by measuring
the additional phase difference that the right-~and
  left-circularly polarized components of the test ray acquire when propagating in the gravitational
field as explained in the end of section~\ref{rot}. The measurement
precision of the phase $\Phi =- \lambda_{\rm test}\Delta$ 
is restricted
by the shot noise. Using classical light, the minimal uncertainty in a
phase estimation cannot exceed the shot noise limit, which is of the
order of magnitude $\delta\Phi\sim \frac{1}{\sqrt{nM}}$, where $n$ is
the number of photons of the light inside the cavity and $M$ the
number of measurements \cite{schottky_1918}.
For a cavity resonator driven by a laser with frequency 
$\omega/(2\pi)$ and power $P_\mathrm{dr}$, we find a number of photons
 $n = P_\mathrm{dr} T_\mathrm{av}/(\hbar \omega)$, where $T_\mathrm{av}$ 
 is the average time a photon spends in the resonator. 
Therefore, the number of measurements that can be performed with
$n$ photons in an experimental time $T_\mathrm{tot}$ is given as $M =
T_\mathrm{tot} / T_\mathrm{av}$, giving
$nM = P_\mathrm{dr} T_\mathrm{tot} /(\hbar \omega)$, 
which is the total number of photons passing the
cavity in time $T_\mathrm{tot}$.    

The measurement precision becomes thus better by increasing the power
of the driving laser and lowering its frequency. For cw-laser beams
with power $P_\mathrm{dr} = 100\,{\rm kW}$
\cite{shcherbakov_2013},\footnote{Of course the power of the driving
  laser cannot be unlimited as the cavity mirrors have to withstand
  the heating due to scattered light. { The finesse $F\sim 10^6$ leads to a
  circulating power in the cavity of the order of $10^{10}\,\rm{W}$,}
  which leads to a necessary size of the beam at the mirrors of the
  order of $1\,\rm{m}$ \cite{Meng:05}. Assuming the transversal setup
  described in section~\ref{cavity}, the waist of the test ray has
  to be smaller than the waist of the source beam and the divergence
  angle of the test ray must be smaller than one radian to
  ensure a complete overlap of the focal regions of the
  source beam and the test ray. We assumed a waist of the source
  beam of the order of $10^{-6}\,\rm{m}$, which implies  
a maximum waist of the test ray of the same order. Furthermore,
the divergence angle of the test ray below one radian implies that
the distance between the mirrors of the test ray has to be of
the order of several meters. The situation for the longitudinal cavity
turns out to be even more challenging. However, the given parameters
serve as an upper limit of what would be possible in the near future.}
for a wavelength of approximately $500\,{\rm nm}$ and a total
experimental time of about two weeks, i.e. $T_\mathrm{tot} \sim
10^6\,{\rm s}$, the minimal standard deviation is given by
$ \delta\Phi\sim 10^{-15}\,{\rm rad}$. Its order of magnitude does not
change when using a squeezed (single mode coherent) state with the
currently maximal squeezing of $15\,{\rm dB}$
\cite{vahlbruch_2016}\footnote{Note that this degree of squeezing has
  only been reached for much a smaller beam power of the order of ${\rm
    mW}$, which would actually lead to a decrease in the sensitivity.} and
analyzing the uncertainty with the corresponding quantum Cram\'er-Rao
bound \cite{pinel_2013}. 

The Cram\'er-Rao bound is a tight bound on the uncertainty of an
unbiased phase-estimation that can in principle be achieved in a
highly idealized situation, where all other noise sources such as
thermal noise, electronic noise, seismic noise etc.~are
neglected. The sensitivity can be increased by using more
than one mode, but without entangling the modes or creating other non-classical
states no gain in sensitivity at fixed total energy is possible
\cite{RevModPhys.90.035006}. 

For a more practical benchmark of current
state-of-the-art measurement precision, consider the  LIGO
observatory. It obtains a sensitivity for length changes of their arms
of the order of $10^{-20}\,{\rm m}$ (strains of the order of
$10^{-23}\,({\rm Hz})^{-1/2}$ on an arm length of the order of
$10^3\,\rm{m}$ \cite{aasi_2013}), which corresponds to a phase
sensitivity of the order of $10^{-11}\,{\rm rad}$ at about $1000\,{\rm
  nm}$ wavelength. 
Another obstacle is that the source-laser
power of $10^{15}\,{\rm W}$ that we considered here can so far only be
reached in very short pulses, which means that
an extension of our analysis to pulsed source beams will be required
when one day substantially larger powers and more sensitive
measurements might become available. We conclude that the angles due
to the gravitational Faraday effect of the order of magnitude
$\Delta\sim 10^{-32}\,{\rm rad}$ cannot be measured with current and
near-future technology. 

%======================================================================================
\section{Summary, Conclusion and outlook}\label{conc}

We analyzed the rotation of polarization for a test ray
propagating in the gravitational field of a laser beam. We
distinguished the non-reciprocal contribution to the rotation due to the
gravitational Faraday effect from the reciprocal
contribution associated with the gravitational optical activity. 
As the rotation angle is equivalent to a phase for
circularly polarized test rays, the precision of the measurement
of the effect investigated in this article is limited by the
shot-noise limit when using classical light. With this analysis we
found that the rotation of polarization of a test ray induced by the
gravitational field of a circularly polarized source laser-beam is too
small to be measured with state-of-the-art technology. The effects
are of fundamental interest, however.

For an infinitely extended (or at least very long) test ray 
propagating parallel to the source beam, we
found that the local rotation picked up by the polarization vector of the test ray is proportional 
to the energy density of the source beam. 
In that case, we concluded that effects are only present 
for an overlap of the test ray and the
source beam's region of highest intensity bounded by its width. Using
the approximation of an infinitely extended source beam, such an
overlap is always present for parallel propagating test rays and
we find {a decay of the integrated rotation angle with the inverse of the distance
to the beamline of the source beam.} In the realistic situation of a
finitely extended source beam, this  { dependence on the distance} remains approximately
valid as long as there is a significant overlap. However, for the
finitely extended source beam, there is no overlap for distances from
the beamline larger than the extension of the beamline  multiplied by
the divergence angle of the source beam. 
Above that limit, we find that the polarization rotation picked up by a 
parallel propagating infinitely extended test ray decreases as a Gaussian with the distance to the beamline of the source beam.
For a finitely extended test ray 
far from the beamline of the source beam, we find that 
the effects decay with the inverse of the third power of the distance
using a multipole expansion. However, a finitely extended test ray begins
and ends in regions with non-vanishing gravitational effect of the source beam.
Hence, the interpretation of the rotation angle is not straight forward. 
To overcome this problem, a physical reference system could be considered that 
extends or is moved from the beginning to the end of the test ray. 

For transversally propagating test rays, the situation is different: The leading order effect decreases with the inverse of the
distance from an finitely extended source beam for an infinitely extended
test ray and with the inverse square for a finitely extended test ray. 
Therefore, of the effects investigated in this article, the rotation of polarization of a
transversal test ray should be the easiest to detect, while we reiterate 
that a detection will not be possible in the near future.
It is interesting to note that the effect remains there also in the geometric optical limit and is
independent of the source beam's helicity. 

Only the gravitational Faraday effect
contributes to the leading order effect for the transversal test ray. 
The gravitational optical activity induces the next to leading order term, 
and it decays one order more strongly with $\chi$ than the gravitational Faraday effect.

It has been shown that for light passing through or being emitted from
a rotating spherical body \cite{skrotsky_1957,balazs_1957} or a
rotating spherical shell \cite{sereno_2004}, one obtains a rotation of
the polarization proportional to the inverse of the square of the
distance to the rotating object. On the other hand, when the light ray
is only passing by these objects or any stationary object, there is no
rotation of polarization
\cite{faraoni_1992,kobzarev_1988,kobzarev_1988}. However, if these
objects are in motion, it has been shown that the polarization is
rotated (for a moving point mass \cite{pen_2017},  for gravitational
lenses \cite{kobzarev_1988,kopeikin_2002},  for a moving Schwarzschild
object \cite{lyutikov_2017},  for moving stars
\cite{plebanski_1960}). As the laser beam, although its spacetime
metric is stationary, consists of an energy-distribution in motion,
our results agree with the literature in the sense that the rotation
of polarization is non-vanishing.

As another interesting fundamental insight, we found that to first
order in the divergence angle $\theta$, the polarization vector of a
parallel counter-propagating test ray rotates, while this is not
the case for a co-propagating test ray. We argue that this
asymmetry is due to the propagation of the source
laser-beam. This is similar to the
deflection of a parallel test ray by the gravitational field of
a laser beam which is non-zero for a counter-propagating ray and
vanishes for a co-propagating ray \cite{tolman}.

The gravitational field of the laser beam depends on its
polarization. This is in agreement with the gravitational field of a
polarized infinitely thin laser beam or pulse derived in
\cite{raetzel_2016} and the gravitational field of a polarized
electromagnetic plane wave presented in
\cite{vanholten_2011}. However, the gravitational field in the models
\cite{raetzel_2016,vanholten_2011} does not depend on the direction of
linear polarization and neither on the helicity of light in the case
of circular polarization. This is in contrast to gravitational
photon-photon scattering in perturbative quantum gravity discussed in
\cite{barker_1967}. In \cite{gaussstrahl}, we showed that the
gravitational field of a laser beam considered as a proper
perturbative solution of Maxwell's equations beyond the short
wavelength approximation does depend on the helicity of the laser
beam. In the present article, we showed that, accordingly, the
polarizations of two light beams couple gravitationally; two
circularly polarized light beams inflict on each other a phase shift
depending on the relation between their helicity. This is
gravitational spin-spin coupling of light (see \cite{mashhoon_2000}
for a general review on gravitational spin-spin coupling). 

Together with frame-dragging and the deflection of a parallel
co-propagating test ray discussed in \cite{gaussstrahl}, the
gravitational Faraday effect and gravitational optical activity are
only visible when the source is treated beyond geometric ray
optics. It can be expected that orbital angular momentum of
light would contribute to the effects mentioned above (see \cite{strohaber_2018}
for an investigation of the gravitational field of light
beams with orbital angular momentum).

%======================================================================================
\section*{acknowledgements}
We thank Marius Oancea for helpful remarks and discussions and Julien Fra\"isse for proofreading the manuscript. DR would like to thank the Humboldt Foundation for supporting his work with their Feodor Lynen Research Fellowship.

%======================================================================================
\appendix

%========================================================
\section{Metric perturbation (from \cite{gaussstrahl})\label{A}}

In this appendix, we give the explicit expressions for the metric
perturbation as derived in \cite{gaussstrahl}. The metric perturbation
is obtained from the electromagnetic field of a circularly polarized
laser beam given in \cite{gaussstrahl}, which is determined by the
vector potential $A_\alpha(\tau,\xi,\chi,\zeta)=\tilde{\mathcal{A}}
v_\alpha(\xi,\chi,\theta \zeta)e^{i\frac{2}{\theta }(\zeta-\tau)}$,
where $\tilde{\mathcal{A}}$ is the amplitude, $v_\alpha = \sum_{n=0}^\infty
\theta^{n}v^{(n)}_{\alpha}$ is the envelop function, whose spatial
components, $a\in\{\xi,\chi,\zeta\}$, are given up to third order in
$\theta$ by 
\begin{eqnarray}\label{eq:circpolenv}
	v^{\lambda (0)}_{a}
	&=& \epsilon^{(0)}_{a}v_0\;,\\
	v^{\lambda (1)}_{a}
	&=& -\epsilon^{(1)}_{a} \frac{i\mu}{2\sqrt{2}}\left(\xi - i\lambda\chi\right)v_0\;,\\
	v^{\lambda (2)}_{a}
	&=&	\frac{\mu}{2}\left(1 - \frac{1}{2}\mu^2\rho^4 \right)v^{\lambda (0)}_{a}\;,\\
	v^{\lambda (3)}_{a}
	&=&	\frac{\mu}{4}\left(4 +  \mu\rho^2 - \mu^2\rho^4 \right)v^{\lambda (1)}_{a}\;,
\end{eqnarray}
where $\mu=1/(1+i\theta\zeta)$, the function $v_0$ is given by
\begin{equation}	
	v_0(\xi,\chi,\theta\zeta) = \mu e^{-\mu\rho^2}\;,\;
\end{equation}
and $\epsilon_{a}^{(0)}=w_0(1,-\lambda i,0)/\sqrt{2}$, $\epsilon^{(1)}_{a}=w_0(0,0,1)$ and $\lambda=\pm 1$ refers to the helicity. Since we work in the Lorenz gauge, the $\tau$-component of the vector potential is given as
\begin{equation}\label{eq:lorenzAtau}
	 A_{\tau} = \frac{i\theta}{2} \partial_{\tau} A_{\tau} =  \frac{i\theta}{2} \left(\partial_\xi A_\xi + \partial_\sigma A_\sigma + \theta\partial_{\theta\zeta} A_{\zeta} \right)  \,.
\end{equation}
The leading order is thus the usual expression for the electromagnetic
field of the Gaussian beam in the paraxial approximation. The higher
orders are corrections to the paraxial approximation. The
corresponding components of the energy-momentum tensor are given as 
$T_{\tau\tau}=\mathcal{E}$, $T_{\tau j}=-S_j/c$ and
$T_{jk}=\sigma_{jk}$ for $j,k\in\{\xi,\chi,\zeta\}$. 
For the vector potential of a circularly polarized laser beam given by
equation~(\ref{eq:circpolenv}), the energy density $\mathcal{E}$, the
Poynting vector $\vec S$ and the stress tensor components
$\sigma_{jk}$ up to third order in $\theta$ are given as 
\begin{eqnarray}
	\mathcal{E}^\lambda 
	&=& \mathcal{E}^{(0)} \Bigg[ 
			1 \\\A&&+ \frac{|\mu|^2\theta^2}{2}\bigg(1 + 	|\mu|^2(2 - (4|\mu|^2 - 3)\rho^2)\rho^2\bigg)\Bigg], \\\
	 S^\lambda_\xi /c
	&=&  \mathcal{E}^{(0)} \theta |\mu|^2\bigg[  (\theta \zeta  \xi +\lambda  \chi ) \\
	\A && - \frac{\theta^2}{4}   \bigg(\lambda\chi - 2|\mu|^2\bigg((2-\rho^2)\theta\zeta\xi + 2(1-\rho^2)\lambda\chi  \\
	\A && + (\theta\zeta\xi + \lambda\chi)(4 + 3\rho^2 - 4|\mu|^2 \rho^2)|\mu|^2\rho^2 \bigg)\bigg)\bigg]\;,
\\
	 S^\lambda_\chi/c
	 &=&	 - \lambda \mathcal{E}^{(0)}  \theta |\mu|^2\bigg[  ( \xi - \theta \zeta \lambda\chi ) \\
	\A && -  \frac{\theta^2}{4}   \bigg(\xi - 2|\mu|^2\bigg(2(1-\rho^2)\xi - (2-\rho^2)\theta\zeta\lambda\chi \\
	\A && + (\xi - \theta\zeta\lambda\chi)(4 + 3\rho^2 - 4|\mu|^2 \rho^2 )|\mu|^2\rho^2 \bigg)\bigg)\bigg]\;,
\\
		S^\lambda_\zeta/c
	&=&	\mathcal{E}^\lambda - \frac{1}{2}\mathcal{E}^{(0)} (\theta\rho|\mu|)^2\;, \\
	 \sigma^\lambda_{\xi\xi}
	&=&	\mathcal{E}^{(0)} \theta^2 |\mu|^4 (\theta \zeta  \xi + \lambda  \chi )^2 \;,\\
	\sigma^\lambda_{\chi\chi}
	&=&	\mathcal{E}^{(0)} \theta^2 |\mu|^4 (  \xi - \theta \zeta \lambda  \chi )^2\;, \\
	\sigma^\lambda_{\xi\chi}
	&=&	 \mathcal{E}^{(0)} \lambda  \theta^2 |\mu|^4  (\theta \zeta  \xi +\lambda  \chi ) (\theta \zeta  \lambda  \chi -\xi )	\;,\\
	\sigma^\lambda_{\xi\zeta}
	&=&  S^\lambda_\xi/c  - \mathcal{E}^{(0)} \frac{\theta^3}{2}   (\theta \zeta  \xi +\lambda  \chi )|\mu|^4\rho^2 \;,\\
	\sigma^\lambda_{\chi\zeta}
	&=&  S^\lambda_\chi/c  +  \lambda \mathcal{E}^{(0)} \frac{\theta^3}{2}  (\xi -  \theta \zeta  \lambda \chi )|\mu|^4\rho^2 \;,\\
	\sigma^\lambda_{\zeta\zeta} 
	&=&  \mathcal{E}^\lambda - \mathcal{E}^{(0)} (\theta\rho|\mu|)^2 \;,
\end{eqnarray}
where $|\mu|^2=1/(1+(\theta\zeta)^2)$ and $\mathcal{E}^{(0)} = \varepsilon_0 w_0^2 E_0^2 |v_0|^2 = 2P_0 |\mu|^2 \mathrm{Exp}(- 2|\mu|^2\rho^2) /(\pi c) $.

%==================================================================================
\subsection{Field equations}

The linearized Einstein equations take the form
\begin{eqnarray}\label{eq:poissonh0}
	\Delta_{2d} h_{\alpha\beta}^{\lambda(0)}
	&=& -\kappa w_0^2\,  t^{\lambda(0)}_{\alpha\beta}\;,\\
	\label{eq:poissonh1}
	\Delta_{2d} h_{\alpha\beta}^{\lambda(1)}
	&=& -\kappa w_0^2 \, t^{\lambda(1)}_{\alpha\beta}\;,\\
	\label{eq:poissonhhigher}\Delta_{2d} h_{\alpha\beta}^{\lambda(n)}
	&=&	-\kappa w_0^2\,  t^{\lambda(n)}_{\alpha\beta} - \partial_{\theta\zeta}^2 h_{\alpha\beta}^{\lambda(n-2)}\;\rm{for}\,\,n > 1\,,
\end{eqnarray}
where $t^{(n)}_{\alpha\beta}$ are the coefficients of the power series expansion of the energy-momentum tensor in orders of $\theta$, i.e.~$T_{\alpha\beta}=\sum_n \theta^n t^{(n)}_{\alpha\beta}$.

%==================================================================================
\subsection{Zeroth order}

The metric perturbation in the leading (zeroth) order of the expansion in the beam divergence is given by \cite{gaussstrahl}
\begin{equation}
h_{\tau\tau}=h_{\zeta\zeta}=-h_{\tau\zeta}= I^{(0)}\;,
\end{equation}
where the function $I^{(0)}$ is given by
\begin{equation}
I^{(0)}=\frac{8GP_0 w_0^2}{ c^5}
\left( \frac{1}{2}{\rm Ei}\left(-2|\mu|^2\rho^2\right) -\log(\rho)\right)\;, 
\end{equation}
where ${\rm Ei}(x)=-\int_{-x}^\infty dt\;\frac{e^{-t}}{t}$ is the exponential integral.

%==================================================================================
\subsection{First order}

The metric perturbation in the first order of the expansion in the beam divergence is given by \cite{gaussstrahl}\\
\begin{eqnarray}
	h^{\lambda(1)}_{ \alpha \beta}
	&=&	\begin{pmatrix}
			0& I_\xi^{\lambda(1)} & I_\chi^{\lambda(1)} & 0\\
			I_\xi^{\lambda(1)} & 0 & 0 & -I_\xi^{\lambda(1)}\\
			I_\chi^{\lambda(1)} & 0 & 0 & -I_\chi^{\lambda(1)}\\
			0 & -I_\xi^{\lambda(1)} & -I_\chi^{\lambda(1)} & 0
		\end{pmatrix}\;,\label{eq:38}
\end{eqnarray}
where the functions $I_{\xi}^\lambda{}^{(1)}$ and $I_{\chi}^\lambda{}^{(1)}$ given by
\begin{align}
I^\lambda_\xi{}^{(1)}
\nonumber&= \frac{1}{4}\left(\theta\zeta\partial_\xi+\lambda \partial_\chi\right)I^{(0)} \\
&= -\frac{2G  P_0 w_0^2 (\theta \zeta  \xi +\lambda  \chi )}{  c^5 \rho^2}  \left(1-e^{-2 |\mu|^2\rho^2}\right)\;,\\
I^\lambda_\chi{}^{(1)}
\nonumber&= -\frac{1}{4}\left(\lambda\partial_\xi -\theta\zeta\partial_\chi\right)I^{(0)}\\
&= \frac{2G  P_0 w_0^2 (\lambda  \xi - \theta \zeta  \chi )}{ c^5 \rho^2}  \left(1-e^{-2 |\mu|^2\rho^2}\right) \;.\label{eq:40}
\end{align}

%==================================================================================
\subsection{Third order}

The only non-zero components of the metric perturbation in the third order of the expansion in the beam divergence are given by 
\begin{widetext}
\begin{align}
	\A h^{\lambda(3)}_{\tau\xi} 
	=&\; -\frac{G  P_0 w_0^2 }{2   c^5 \rho^2} \Bigg( (4\theta\zeta\xi + 3\lambda\chi) + \Bigg(-(4\theta\zeta\xi + 3\lambda\chi)-2\rho^2(3\theta\zeta\xi + 2\lambda\chi)|\mu|^2\\
	 &\; - 2\rho^2(-2 + 3\rho^2)(\theta\zeta\xi + \lambda\chi)|\mu|^4 + 8\rho^4(\theta\zeta\xi + \lambda\chi) |\mu|^6 \Bigg) e^{-2 |\mu|^2\rho^2}\Bigg)\;,\\
	\A h^{\lambda(3)}_{\tau\chi} =&\; -\frac{G  P_0 w_0^2 }{2   c^5 \rho^2} \Bigg( (4\theta\zeta\chi - 3\lambda\xi) + \Bigg(-(4\theta\zeta\chi - 3\lambda\xi) - 2\rho^2(3\theta\zeta\chi - 2\lambda\xi)|\mu|^2 \\
	&\; - 2\rho^2(-2 + 3\rho^2)(\theta\zeta\chi - \lambda\xi)|\mu|^4 + 8\rho^4(\theta\zeta\chi - \lambda\xi) |\mu|^6 \Bigg) e^{-2 |\mu|^2\rho^2}\Bigg)\;,\\
	\A h^{\lambda(3)}_{\zeta\xi} =&\; \frac{G  P_0 w_0^2 }{2   c^5 \rho^2} \Bigg( (2\theta\zeta\xi + \lambda\chi) + \Bigg(-(2\theta\zeta\xi + \lambda\chi)-2\rho^2(2\theta\zeta\xi + \lambda\chi)|\mu|^2 \\
	&\; - 2\rho^2(-2 + 3\rho^2)(\theta\zeta\xi + \lambda\chi)|\mu|^4 + 8\rho^4(\theta\zeta\xi + \lambda\chi) |\mu|^6 \Bigg) e^{-2 |\mu|^2\rho^2}\Bigg)\;,\\
	\A h^{\lambda(3)}_{\zeta\chi} =&\; \frac{G  P_0 w_0^2 }{2   c^5 \rho^2} \Bigg( (2\theta\zeta\chi - \lambda\xi) + \Bigg(-(2\theta\zeta\chi - \lambda\xi) - 2\rho^2(2\theta\zeta\chi - \lambda\xi)|\mu|^2 \\
	&\; - 2\rho^2(-2 + 3\rho^2)(\theta\zeta\chi - \lambda\xi)|\mu|^4 + 8\rho^4(\theta\zeta\chi - \lambda\xi) |\mu|^6 \Bigg) e^{-2 |\mu|^2\rho^2}\Bigg)\,.
\end{align}
\end{widetext}

%===========================================================
\section{Another approach to determine the rotation of polarization\label{B} (as described in \cite{pen})}

Another result for the rotation of the polarization was obtained in \cite{pen}, where the polarization vector is parallel transported through the gravitational field, again starting and ending in flat spacetime. 
The angle of rotation in the $\alpha\beta$-plane is given by
\begin{equation}
\tilde\Delta_{\alpha\beta}
= 
\int_{-\infty}^\infty d\tilde\tau\; \dot\gamma^\gamma \Gamma^\delta_{\alpha\gamma}g_{\beta\delta}\;,
\label{rotpen}
\end{equation}
where $\tilde\tau$ is the parameter parametrizing the geodesic $\gamma$.
It is obtained as follows: The polarization vector 
$\omega^\alpha$ is parallel transported if \begin{equation} 
\dot\gamma^\alpha\partial_\alpha \omega^\gamma +\dot\gamma^\alpha \omega^\beta\Gamma^\gamma_{\alpha\beta}
=0\;.
\end{equation}
Integrating along the geodesic $\gamma$, the change of polarization is given by
\begin{equation}
\delta \omega^\gamma
= \int_{-\infty}^\infty d\tau\; \dot\gamma^\alpha\partial_\alpha \omega^\gamma
= -\int_{-\infty}^\infty d\tau\;
\dot\gamma^\alpha \omega^\beta \Gamma^\gamma_{\alpha\beta}\;.
\end{equation}
From the change of polarization, the angle of rotation in the plane $\beta\gamma$ is obtained by writing
\begin{equation}
(\omega+\delta\omega)^\gamma
=\big(
g^\gamma_{\;\;\beta} +\tilde\Delta^\gamma_{\;\;\beta}
\big)\omega^\beta\;,
\end{equation}
which has the form of an infinitesimal rotation. The rotation angle is
given by (\ref{rotpen}). This result is coordinate-invariant if the
metric perturbation vanishes far away from the source of the
gravitational field. This is not the case for the laser beam. However,
in some cases the result can be applied, as we will explain. Also,
(\ref{rotpen}) describes a four-dimensional rotation. If the test
light-ray is deflected by the laser beam (as for the parallel
counter-propagating and the transversal light ray), one has to be
careful when applying this formula, as the ray-transversal plane tilts
when the light ray is deflected. 
In our case, the formula can be applied. Indeed, it leads to the same results as we obtain with equation~(\ref{eq:rotpleb}): For the parallel co-~and parallel conter-propagating light rays, one obtains (to third and first order in the expansion in $\theta$, respectively)
\begin{align}
\tilde\Delta^+_{\xi\chi}\A
=& -\frac{\theta^2}{2w_0^2}\int_{-\infty}^\infty d(\theta\zeta)\, \Big(\partial_\chi \left(h^{(3)}_{\xi\zeta}+h^{(3)}_{\tau\xi}\right)\\
& -\partial_\xi\left( h^{(3)}_{\chi\zeta}+h^{(3)}_{\tau\chi}\right)-\partial_{\theta\zeta} h^{(2)}_{\xi\chi} \Big)\;,\\
\tilde\Delta^-_{\chi\xi}\A
=&-\frac{1}{2w_0^2}
\int_{-\infty}^\infty d(\theta\zeta)
\bigg(
\partial_\chi\left(h_{\xi\zeta}^{(1)}-h_{\xi\tau}^{(1)}\right)\\
&-\partial_\xi\left(
h_{\chi\zeta}^{(1)}-h_{\chi\tau}^{(1)}
\right)
\bigg)\;.
\end{align}
The last term of the integrand in the above equation for $\tilde\Delta^+_{\xi\chi}$ vanishes when
integrating from $\zeta=-\infty$ to $\zeta=\infty$, as in our case
$h_{\xi\chi}(\infty)=h_{\xi\chi}(-\infty)$. Therefore, we see that $\tilde\Delta^+_{\xi\chi}=\Delta_+$ and $\tilde\Delta^-_{\chi\xi}=\Delta_-$. The same is the case for the transversally propagating light rays: We find (up to the first order in the expansion in $\theta$)
\begin{align}
\tilde\Delta_{\chi\zeta}^{t^+}
&= \frac{1}{2w_0^2}\int_{-\infty}^\infty d\xi
\bigg(
\partial_\chi h_{\tau\zeta}^{(0)} 
- \theta\partial_\chi h_{\xi\tau}^{(1)}
+\theta\partial_\xi h_{\chi\zeta}^{(1)}
\bigg)
\;,
\\
\tilde\Delta_{\zeta\chi}^{t^-}
&=\frac{1}{2w_0^2}\int_{-\infty}^\infty d\xi
\bigg(
-\partial_\chi h_{\tau\zeta}^{(0)} 
- \theta\partial_\chi h_{\xi\tau}^{(1)}
-\theta\partial_\xi h_{\chi\zeta}^{(1)}
\bigg)
\;.
\end{align}
As $h^{(1)}_{\chi\zeta}(\xi=\infty)=h^{(1)}_{\chi\zeta}(\xi=-\infty)$, we obtain $\tilde\Delta_{\chi\zeta}^{t^+}=\Delta_{t^+}$ and $\tilde\Delta_{\zeta\chi}^{t^-}=\Delta_{t^-}$.

%===========================================================
\section{ Derivation for finitely extended source and test beams\label{C}}

Starting from the solution in equation (\ref{eq:primeli}) for the
linearized Einstein equations, we find with equation (\ref{pm}), using
the identity $\partial_{x^a}\frac{1}{|\vec{x}-\vec{x}'|} =
- \partial_{x^{a\prime}}\frac{1}{|\vec{x}-\vec{x}'|} $, and partial 
integration (the energy-momentum tensor vanishes at infinity) 
\begin{align}\label{eq:DeltapmT} 
\A \Delta_{\pm}
&= -\frac{2G}{c^4}\int_{A}^B d\zeta \int_{-\infty}^\infty d\xi' d\chi' d\zeta' \, \frac{1}{|\vec{x}-\vec{x}'|} \\\A
&	\Big(
\theta\Big(\partial_{\chi'}\left(t^{(1)}_{\xi\zeta} \pm t^{(1)}_{\tau\xi}\right) -
  \partial_{\xi'}\left(t^{(1)}_{\chi\zeta} \pm t^{(1)}_{\tau\chi}\right)\Big)\\
 & +
\theta^3\Big(\partial_{\chi'}\left(t^{(3)}_{\xi\zeta} \pm t^{(3)}_{\tau\xi}\right)
 -\partial_{\xi'}\left(t^{(3)}_{\chi\zeta} \pm t^{(3)}_{\tau\chi} \right)\Big)
\Big)\;.
\end{align}
The energy-momentum tensor of the finitely extended beam is given by
multiplying 
the expressions in appendix A for the infinitely extended beam 
with the Heaviside functions
$\Theta(\zeta-\alpha(\rho))$ and $\Theta(\beta(\rho)-\zeta)$, where
$\alpha(\rho)$ and $\beta(\rho)$ describe the $\zeta$-coordinate of
the source beam's emitter and absorber, respectively. This
truncation of the energy-momentum tensor leads to a violation of the
continuity equation of general relativity, which in our case
means neglecting recoil on emitter and absorber. This corresponds to
energy and momentum being inserted  into the system and
dissipated from it, respectively, and can lead to apparent effects
close to emitter and absorber that may not be present in
practice. The best approximation of reality by our model of the
finitely extended beam will be achieved for points far from emitter
and absorber but close to the beamline (see also
\cite{Raetzel:2015xbc} for a detailed analysis of a similar
situation).  

When the surfaces of emitter and absorber are considered to match the
phase fronts of the beam, they are curved and, therefore, depend on
$\rho$. This dependence is of second order in $\theta$. The
derivatives in equation (\ref{eq:DeltapmT}) lead to Dirac delta
functions $\alpha'(\rho)\delta(\zeta-\alpha(\rho))$ and
$\beta'(\rho)\delta(\beta(\rho)-\zeta)$,  
and hence to evaluation of
the integrand at the surfaces of emitter and absorber, respectively,
integrated over the transversal directions. For each term in equation
(\ref{eq:DeltapmT}), this 
contributes even higher order terms. In the
following, we restrict our considerations to the leading order only
(to first order for $\Delta_-$ and to third order for
$\Delta_+$). Therefore, the contributions of the curved surfaces of
emitter and absorber can be neglected and we set $\alpha$ and $\beta$
to be constants.
From the expressions given in appendix~\ref{A} for the energy-momentum
tensor, one sees that $t^{(1)}_{\xi\zeta} = -t^{(1)}_{\tau\xi}$ and
$t^{(1)}_{\chi\zeta} = -t^{(1)}_{\tau\chi}$. The derivatives appearing in the expression for $\Delta_\pm$ of the first order terms are given by {\small
\begin{eqnarray}
	\partial_\chi t_{\xi\zeta}^{(1)}	&=& \frac{2P_0}{\pi c}|\mu|^4 \Big(-4\chi|\mu|^2(\theta\zeta\xi+\lambda\chi)+\lambda\Big)e^{-2|\mu|^2\rho^2}\,,\\
	\partial_\xi t_{\chi\zeta}^{(1)}&=& \frac{2P_0}{\pi c}|\mu|^4 \Big(-4\xi|\mu|^2(\theta\zeta\chi - \lambda\xi)-\lambda\Big)e^{-2|\mu|^2\rho^2}\,,
\end{eqnarray}}
and the derivatives of the third order terms are found to be {\small
\begin{eqnarray}
\A	\partial_\chi \Big( t_{\xi\zeta}^{(3)} + t_{\tau\xi}^{(3)}\Big) 
	&=&  -\frac{P_0}{\pi c}|\mu|^6 \rho^2 e^{-2|\mu|^2\rho^2} \Big(\lambda \\
 && + (-4|\mu|^2+2/\rho^2)\chi(\theta\zeta\xi + \lambda\chi) \Big)\,,\\\A
		\partial_\xi \Big( t_{\chi\zeta}^{(3)} + t_{\tau\chi}^{(3)}\Big) 
	&=& -\frac{P_0}{\pi c}|\mu|^6 \rho^2 e^{-2|\mu|^2\rho^2}\Big( - \lambda \\
 && + (-4|\mu|^2+2/\rho^2)\xi(\theta\zeta\chi - \lambda\xi)\Big)\,.
\end{eqnarray}}
Considering only the leading order terms in $\theta$, we obtain for
the rotation angles of the parallel co-~and the parallel
counter-propagating test rays
\begin{eqnarray}
	\A \Delta_- &=& - \frac{8GP_0}{c^5}\frac{2\lambda\theta}{\pi}  \int_{-\infty}^\infty d\xi' d\chi' \int_\alpha^\beta d\zeta' K(\xi',\chi',\zeta') \label{58}\\
	&&|\mu(\zeta')|^4 (1-2|\mu(\zeta')|^2\rho'^2)e^{-2|\mu(\zeta')|^2\rho'^2}\,,\\
	\A \Delta_+ &=& \frac{8GP_0}{c^5}\frac{\lambda\theta^3}{\pi}  \int_{-\infty}^\infty d\xi' d\chi' \int_\alpha^\beta d\zeta' K(\xi',\chi',\zeta')\label{59} \\
	&&|\mu(\zeta')|^6\rho'^2 (1-|\mu(\zeta')|^2\rho'^2)e^{-2|\mu(\zeta')|^2\rho'^2}\,,
\end{eqnarray}
where $|\mu(\zeta')|^2=1/(1+(\theta\zeta')^2)$ and
\begin{equation}
	K(\xi',\chi',\zeta') = \log\left(\frac{B-\zeta' + (\rho''^2 + (B-\zeta')^2)^{1/2}}{A-\zeta' + (\rho''^2 + (A-\zeta')^2)^{1/2}}\right)\,,
\end{equation}
with $\rho''=\sqrt{(\xi'-\xi)^2+(\chi-\chi')^2}$. 

For the transversal test ray, we find along the same lines (neglecting again the effect of the curved surfaces of emitter and absorber as they are at least of second order in $\theta$), using equation (\ref{eq:tpmgen}),
\begin{align}\A
\Delta_{t\pm}
=& \frac{2G}{c^4}\int_{A}^B d\xi \int_{-\infty}^\infty d\xi' d\chi' d\zeta' \, \frac{1}{|\vec{x}-\vec{x}'|} \\
&	\partial_{\chi'}\left(\pm t^{(0)}_{\tau\zeta} + \theta t^{(1)}_{\xi\zeta}\right)\,.
\end{align}
From the expressions for the energy-momentum tensor in appendix A, we find that the derivatives in the above equation are given by 
\begin{eqnarray}
	 \partial_\chi t_{\tau\zeta}^{(0)} 
	&=& \frac{8P_0}{\pi c}|\mu|^4\chi e^{-2|\mu|^2\rho^2}	\,,\\
 \A \partial_\chi t_{\xi\zeta}^{(1)} &=& \frac{2P_0}{\pi c}|\mu|^4(\lambda(1-4|\mu|^2\chi^2)\\
 && -4\theta\zeta\xi\chi|\mu|^2)e^{-2|\mu|^2\rho^2}
\end{eqnarray}
which leads to the rotation angle for the transversal test ray
\begin{eqnarray}\label{eq:Deltatnum}
	\A \Delta_{t\pm} &=&  \frac{8GP_0}{c^5}\frac{1}{2\pi}  \int_{-\infty}^\infty d\xi' d\chi' \int_\alpha^\beta d\zeta' K_t(\xi',\chi',\zeta') \\
	\A &&|\mu(\zeta')|^4 \Big(\pm 4\chi' + \theta (\lambda(1-4|\mu(\zeta')|^2\chi'^2) \\
	&& -4\theta\zeta'\xi'\chi'|\mu(\zeta')|^2)\Big) e^{-2|\mu(\zeta')|^2\rho'^2}\,,
\end{eqnarray}
where the function $K_t$ is given by
\begin{align}
	& \A K_t(\xi',\chi',\zeta') \\
	& = \log\left(\frac{B-\xi' + (\chi''^2 + (\zeta-\zeta')^2 + (B-\xi')^2)^{1/2}}{A-\xi' + (\chi''^2 + (\zeta-\zeta')^2 + (A-\xi')^2)^{1/2}}\right)\,,
\end{align} 
where $\chi''=\chi'-\chi$. 

For the numerical analysis, we transform
the found expressions for the rotation angles into the cylindrical
coordinates $\big(\rho',\phi',\zeta'\big)$ with $\phi'=\arccos(\xi'/\rho')$ or
$\big(\rho'',\phi'',\zeta'\big)$ with $\rho''=\sqrt{\xi'^2 + \chi''^2}$ and $\phi''=\arccos(\xi'/\rho'')$.

%===========================================================
\section{Derivation for infinitely extended source and test beams\label{D}}

For the parallel test rays, we obtain from equation (\ref{eq:rotpleb}) and { $t_{0,\pm} = \dot\gamma_\pm(\tau_0) =(1,0,0,\pm (1 - f^\pm))$} 
\begin{align}\label{pmapp} 
\Delta_{\pm} =& \frac{1}{2w_0^2} \int_{-\infty}^\infty d \tau\;
 t_0^a \epsilon_{abc} \partial_b h_{c\alpha}(\varrho_\perp + \tau t_0 ) t_0^\alpha
 \\\A 
  = & {\frac{1}{2w_0^2} \int_{-\infty}^\infty d \tau\;
\epsilon_{\zeta bc} \partial_b \left(h_{c\zeta}(\xi,\chi,\pm \tau)  \pm h_{c\tau}(\xi,\chi,\pm \tau)\right) }
\\
 \A = & - \frac{1}{2w_0^2} \int_{-\infty}^\infty d \zeta\;
\Big( \partial_\chi(h_{\xi\zeta} \pm h_{\xi\tau}) - \partial_\xi(h_{\chi\zeta} \pm h_{\chi\tau})\Big)\,.
\end{align}
The rotation angle for the parallel counter-propagating test ray is thus given by (considering the leading order only)
\begin{align}\A
\Delta_{-} =& { -\frac{\theta}{2w_0^2}\int_{-\infty}^\infty d\zeta}
\Big(\partial_\chi\left(h^{(1)}_{\xi\zeta} - h^{(1)}_{\tau\xi}\right)\\
 &-
  \partial_\xi\left(h^{(1)}_{\chi\zeta} - h^{(1)}_{\tau\chi}\right)\Big)\;.
\end{align}
From the expressions for the metric perturbation in appendix A, 
we see that $h^{(1)}_{\xi\zeta}=-h^{(1)}_{\tau\xi}$, $h^{(1)}_{\chi\zeta}=-h^{(1)}_{\tau\chi}$. 
For the derivatives in the above expression, we find
\begin{align}
 \partial_\chi h^{(1)}_{\xi\zeta} - \partial_\xi h^{(1)}_{\chi\zeta} 
 &=  \frac{8GP_0w_0^2}{c^5}\lambda |\mu|^2 e^{-2|\mu|^2\rho^2}\;,
\end{align}
which leads to the rotation angle for the parallel counter-propagating test ray
\begin{align}
\Delta_{-} =& { -\lambda\frac{8GP_0\theta}{c^5}\int_{-\infty}^\infty d\zeta}\,|\mu|^2 e^{-2|\mu|^2\rho^2}\,.
\end{align}
Along the same lines, we find in leading order
\begin{align}
\Delta_{+} =&\A { -\frac{\theta^3}{2w_0^2}\int_{-\infty}^\infty d\zeta}\,
\Big(\partial_\chi\left(h^{(3)}_{\xi\zeta} + h^{(3)}_{\tau\xi}\right)\\
 &-
  \partial_\xi\left(h^{(3)}_{\chi\zeta} + h^{(3)}_{\tau\chi}\right)\Big)\;.
\end{align}
From the expressions for the metric perturbation in appendix A, one
finds for the derivatives in the above expression 
\begin{align}\A
\A & \partial_\chi\left(h^{(3)}_{\xi\zeta} + h^{(3)}_{\tau\xi}\right) - \partial_\xi\left(h^{(3)}_{\chi\zeta} + h^{(3)}_{\tau\chi}\right)\\
= & -\lambda \frac{2GP_0 w_0^2}{c^5}|\mu|^2(1+2\rho^2|\mu|^2)e^{-2|\mu|^2\rho^2}\;.
\end{align}
Then, the rotation angle for the parallel co-propagating light ray is given by
\begin{align}
\Delta_{+} =& { \lambda\frac{ GP_0 \theta^3}{c^5}\int_{-\infty}^\infty d\zeta}\,|\mu|^2(1+2\rho^2|\mu|^2)e^{-2|\mu|^2\rho^2}\,.
\end{align}
For the transversal test ray, we obtain from equation (\ref{eq:rotpleb}) and $\dot\gamma_\pm=(1,\pm 1,0,0)$
\begin{align}\A
\Delta_{t^\pm} =& \frac{1}{2w_0^2} \int_{-\infty}^\infty d \tau\;
 t_0^a \epsilon_{abc} \partial_b h_{c\alpha}(\tau,\varrho_\perp + \tau t_0 ) t_0^\alpha \\
 \A = & \pm \frac{1}{2w_0^2} \int_{-\infty}^\infty d\xi \left(\partial_\chi h_{\tau\zeta} - \theta\partial_{\theta\zeta} h_{\tau\chi}\right) \\
 \label{eq:tpmgen} & + \frac{1}{2w_0^2} \int_{-\infty}^\infty d\xi \left(\partial_\chi h_{\xi\zeta} - \theta\partial_{\theta\zeta} h_{\xi\chi}\right)\,,
\end{align}
Considering the terms up to first order in $\theta$, it is given by
\begin{align}
\Delta_{t^\pm} = & \pm\frac{1}{2w_0^2}\int_{-\infty}^\infty d\xi \,\partial_\chi h_{\tau\zeta}^{(0)}
+\frac{\theta}{2w_0^2}\int_{-\infty}^\infty d\xi\, \partial_\chi h_{\xi\zeta}^{(1)}\,.
\end{align}
From the expressions for the metric perturbation in appendix A, we obtain for the derivatives appearing in the above expression
\begin{align}
	 \partial_\chi h^{(0)}_{\tau\zeta} 
	  = & \frac{8GP_0w_0^2}{c^5} \frac{\chi}{\rho^2}\Big(1-e^{-2|\mu|^2\rho^2}\Big) 
\;,\\\label{abv}
	 \partial_\chi h^{(1)}_{\xi\zeta} = & -\frac{1}{4}(\theta\zeta\partial_\chi\partial_\xi + \lambda\partial_\chi^2)I^{(0)}\,.
\end{align}
The first term in equation~(\ref{abv}) leads to an integration over a derivative, which vanishes,
\begin{equation}
	\int_{-\infty}^\infty d\xi \,\partial_\chi \partial_\xi I^{(0)} = \left.\partial_\chi I^{(0)}\right|_{\xi=-\infty}^{\xi=\infty} = 0\,.
\end{equation}
Then, we obtain for the rotation angle for the transversal test ray
\begin{align}
\A \Delta_{t^\pm} = & \pm \frac{4\pi G P_0 }{c^5}\;{\rm erf}\left(\sqrt{2}|\mu|\chi\right)\\
& +\lambda\frac{2\sqrt{2\pi}GP_0\theta}{c^5}\,|\mu| e^{-2|\mu|^2\chi^2} \;.
\end{align}

%===========================================================
\section{Derivation for finitely extended source beams and infinitely extended test rays\label{fininf}}
For an infinitely extended test ray and a finitely extended source beam,
we obtain
\begin{eqnarray}
\A \Delta_{-} \A  &=& -\frac{2G}{c^4}\partial_{\chi} \lim_{B\rightarrow \infty} \int_{-B}^B d\zeta \int_{-\infty}^\infty d\xi'd\chi'd\zeta' \\
 \A &&  \, \frac{1}{|\vec{x}-\vec{x}'|} \Big(t_{\xi\zeta}(\xi',\chi',\zeta') - t_{\tau\xi}(\xi',\chi',\zeta')\Big) \\
\A  &&	+ \frac{2G}{c^4}\partial_{\xi} \lim_{B\rightarrow \infty} \int_{-B}^B d\zeta \int_{-\infty}^\infty d\xi'd\chi'd\zeta' \\
&&\A  \, \frac{1}{|\vec{x}-\vec{x}'|}  \Big(t_{\chi\zeta}(\xi',\chi',\zeta') - t_{\tau\chi}(\xi',\chi',\zeta')\Big)\\
\A &=& -\frac{4G\theta}{c^4}\partial_{\chi}  \int_\alpha^\beta d\zeta'  \int_0^{\rho_0(\zeta')} d\rho'\, \rho'\int_{0}^{2\pi}d\phi' \\
\A &&   \, \lim_{B\rightarrow \infty} K_B(\xi,\chi,\zeta,\rho',\phi',\zeta') t^{(1)}_{\xi\zeta}(\rho',\phi',\zeta')  \\
\A &&	+ \frac{4G\theta}{c^4}\partial_{\xi} \int_\alpha^\beta d\zeta' \int_0^{\rho_0(\zeta')} d\rho'\, \rho'\int_{0}^{2\pi}d\phi' \\
 &&   \, \lim_{B\rightarrow \infty} K_B(\xi,\chi,\zeta,\rho',\phi',\zeta')  t^{(1)}_{\chi\zeta}(\rho',\phi',\zeta')\,,
\end{eqnarray}
where cylindrical coordinates $\rho'=\sqrt{\xi'^2 + \chi'^2}$ and $\phi'=\arctan(\chi'/\xi')$ are used and the function $K_B$ is given by
\begin{eqnarray}
	\A && K_B(\xi,\chi,\zeta,\rho',\phi',\zeta') \\
	&=& \log\left(\frac{B-\zeta' + (\rho''^2 + (B-\zeta')^2)^{1/2}}{-B -\zeta' + (\rho''^2 + (B+\zeta')^2)^{1/2}}\right)\,,
\end{eqnarray}
where
$
	 \rho''^2 = (\xi'-\xi)^2 + (\chi'-\chi)^2 
	= \rho'^2 + \rho^2 - 2\rho'\rho \cos(\phi-\phi')$,
and $\rho_0(\zeta')=\rho_0/|\mu(\zeta')|$ is the finite transversal extension of the beam that is related to the width of emitter and absorber and $\rho_0$ is a constant.
For $\beta/B \ll 1$, $-\alpha/B \ll 1$ and $\rho_0(\zeta')/B \ll 1$ for all $\zeta'\in [\alpha,\beta]$, we obtain 
\begin{eqnarray}
	\A && K_B(\rho',\phi',\zeta')\\
	\A &=& \log\left(\frac{B-\zeta' + (\rho''^2 + B^2(1 - \zeta'/B)^2)^{1/2}}{-B -\zeta' + (\rho''^2 + B^2(1 + \zeta'/B)^2)^{1/2}}\right) \\
	\A & \approx & \log\left(\frac{2(B-\zeta') + \rho''^2/(2B(1 - \zeta'/B))}{\rho''^2/(2B(1 + \zeta'/B))}\right) \\
	\A & = & \log\left(\frac{1 + \zeta'/B}{1 - \zeta'/B} + \frac{4B^2}{\rho''^2}(1 - (\zeta'/B)^2)\right) \\
	&\approx & \log\left(4\frac{B^2}{\rho''^2}\right)\,.\label{abd}
\end{eqnarray}
In order to evaluate the expression for $\Delta_-$, one needs to take derivatives of the function $K_B$. One finds
\begin{eqnarray}
	&& \partial_\chi \log\left(4\frac{B^2}{\rho''^2}\right) t^{(1)}_{\xi\zeta} - \partial_\xi \log\left(4\frac{B^2}{\rho''^2}\right)t^{(1)}_{\chi\zeta} \\
\A &=& \frac{2P_0}{\pi c}|\mu'|^4 e^{-2|\mu'|^2\rho'^2} \Big(\lambda \rho'\partial_{\rho'} + \theta\zeta'\partial_{\phi'}\Big) \log\big(\rho''^2\big)\,.
\end{eqnarray}
Therefore, one finds for the following expression appearing in the expression for $\Delta_-$, 
\begin{align}
\A & \partial_\chi \int_0^{\rho_0(\zeta')} d\rho'\, \rho' \int_{0}^{2\pi}d\phi'  \, \lim_{B\rightarrow \infty}  K_B(\rho',\phi',\zeta') t^{(1)}_{\xi\zeta}  \\
\A & - \partial_\xi \int_0^{\rho_0(\zeta')} d\rho'\, \rho' \int_{0}^{2\pi}d\phi'  \, \lim_{B\rightarrow \infty}  K_B(\rho',\phi',\zeta') t^{(1)}_{\chi\zeta}\\
& =\A \frac{2P_0}{\pi c} |\mu'|^4 \int_0^{\rho_0(\zeta')} d\rho'\, \rho' e^{-2|\mu'|^2\rho'^2}  \\
& \int_{0}^{2\pi}d\phi'  \Big(\lambda\rho'\partial_{\rho'} + \theta\zeta'\partial_{\phi'}\Big) \log\big(\rho''^2\big) \,.
\end{align}
The term containing the $\phi'$-derivative vanishes under the integral. With
\begin{eqnarray}
	\A && \rho' \partial_{\rho'} \int_{0}^{2\pi}d\phi'  \log\big(\rho'^2 + \rho^2 - 2\rho\rho'\cos(\phi'-\phi)\big)\\
	&=& 2\pi \rho' \partial_{\rho'}\left\{\begin{array}{cc}
	\log\Big(\rho'^2\Big) & \rm{for}\quad \rho \le \rho' \\\A
	\log\Big(\rho^2\Big) & \rm{for}\quad \rho > \rho' 
	\end{array}\right\}\\
	&=& 4\pi \left\{\begin{array}{cc}
	1 & \rm{for}\quad \rho \le \rho'\\
	0 & \rm{for}\quad \rho > \rho'
	\end{array}\right\} = 4\pi \Theta(\rho'-\rho)\;,
\end{eqnarray}
we obtain 
\begin{eqnarray}
 && \frac{2P_0\lambda}{\pi c}|\mu'|^4\int_0^{\rho_0(\zeta')} d\rho'\, \rho' \int_{0}^{2\pi}d\phi' \, \\
\A && e^{-2|\mu'|^2\rho'^2} \rho'\partial_{\rho'} \log\big(\rho''^2\big)\\
\A &=& -\frac{2P_0\lambda}{c}|\mu'|^2\int_0^{\rho_0(\zeta')} d\rho'\, \Theta(\rho'-\rho) \partial_{\rho'}e^{-2|\mu'|^2\rho'^2}\\
\A &=& -\frac{2P_0\lambda}{c}|\mu'|^2 \left\{\begin{array}{cc}
	\int_\rho^{\rho_0(\zeta')} d\rho'\, \partial_{\rho'}e^{-2|\mu'|^2\rho'^2} &:\, \rho \le \rho_0(\zeta')\\
\A	0 &:\, \rho > \rho_0(\zeta')
	\end{array}\right\} \\
\A &=& \frac{2P_0\lambda}{c}|\mu'|^2  \Theta(\rho_0(\zeta') - \rho)\Big(e^{-2|\mu'|^2\rho^2} - e^{-2|\mu'|^2\rho_0^2(\zeta')}\Big)\,.
\end{eqnarray}
Finally, we obtain for the rotation of polarization for the parallel counter-propagating test ray
\begin{eqnarray}\label{kuchenfin}
\Delta_{-} &=& { -\lambda\frac{8GP_0\theta}{c^5} \int_\alpha^\beta d\zeta'\,} \\
  \A  && \Theta(\rho_0 - |\mu'|\rho)|\mu'|^2 \Big(e^{-2|\mu'|^2\rho^2} - e^{-2\rho_0^2}\Big)\;,
\end{eqnarray}
which leads to equation (\ref{kuchen}) for $\rho_0\rightarrow \infty$.
We see that $\Delta_-$ vanishes if there is no overlap with the beam, i.e.~if $\rho > \rho_0(\alpha)$ and $\rho > \rho_0(\beta)$. For large $\rho$, there is only an overlap for large $\zeta'$ for which $\rho_0(\zeta')\approx \rho_0 \theta\zeta'$ and $|\mu'|=|\theta\zeta'|^{-1}$. 
Evaluating the integral, we find
\begin{widetext}
\begin{eqnarray}
 \A \Delta_{-} 
   \A &=& \lambda\frac{8GP_0}{c^5\rho} \Bigg[\Theta(-\theta\alpha -\rho/\rho_0)  \Bigg( \frac{\sqrt{\pi}}{2\sqrt{2}}\Bigg( \rm{erf}\left(-\frac{\sqrt{2}\rho}{\theta\alpha}\right) - \rm{erf}\left(\sqrt{2}\rho_0\right)\Bigg) - e^{-2\rho_0^2} \Bigg(- \frac{\rho}{\theta\alpha} - \rho_0\Bigg)\Bigg)\\
    && + \Theta(\theta\beta -\rho/\rho_0) \Bigg( \frac{\sqrt{\pi}}{2\sqrt{2}}\Bigg( \rm{erf}\left(\frac{\sqrt{2}\rho}{\theta\beta}\right) - \rm{erf}\left(\sqrt{2}\rho_0\right)\Bigg) - e^{-2\rho_0^2} \Bigg(\frac{\rho}{\theta\beta} - \rho_0\Bigg)\Bigg)\Bigg]\,.
\end{eqnarray}
\end{widetext}
For $\rho_0\rightarrow\infty$, we obtain
\begin{eqnarray}
\A  \Delta_{-} &= & -\lambda\frac{4GP_0}{c^5\rho} \frac{\sqrt{\pi}}{\sqrt{2}}\\
   &&\Bigg( \rm{erfc}\left(\frac{\sqrt{2}\rho}{\theta\beta}\right) + \rm{erfc}\left(\frac{\sqrt{2}\rho}{\theta|\alpha|}\right)\Bigg) \,,
\end{eqnarray}
where erfc is the complementary error function. For $\rho \gg \theta\beta$ and $\rho \gg -\theta\alpha$, using the asymptotic expansion of the complementary error function, we obtain {\small
\begin{eqnarray}\label{eq:decay}
  \Delta_{-} &\approx & -\lambda\frac{2GP_0\theta}{c^5\rho^2} \Bigg(\beta e^{-2(\rho/\theta\beta)^2} + |\alpha| e^{-2(\rho/\theta\alpha)^2} \Bigg) \,.
\end{eqnarray}}

For $\Delta_+$, it follows from equation (\ref{eq:DeltapmT}) that in leading order (third order in $\theta$), the rotation of polarization for the parallel co-propagating light ray is given by
\begin{eqnarray}
 \Delta_{+}  &=& -\frac{2G}{c^4}\partial_{\chi} \lim_{B\rightarrow \infty} \int_{-B}^B d\zeta  \\
 \A && \int_{-\infty}^\infty d\xi'd\chi'd\zeta' \, \frac{1}{|\vec{x}-\vec{x}'|} \left(t_{\xi\zeta} + t_{\tau\xi}\right) \\
\A  &&	+ \frac{2G}{c^4}\partial_{\xi} \lim_{B\rightarrow \infty} \int_{-B}^B d\zeta  \\
&&\A \int_{-\infty}^\infty d\xi'd\chi'd\zeta' \, \frac{1}{|\vec{x}-\vec{x}'|}  \left(t_{\chi\zeta} + t_{\tau\chi}\right)\\
\A &=& -\frac{2G\theta^3}{c^4}\partial_{\chi}  \int_\alpha^\beta d\zeta'  \int_0^{\rho_0(\zeta')} d\rho'\, \rho' \\
\A && \int_{0}^{2\pi}d\phi'  \, \lim_{B\rightarrow \infty} K_B(\rho',\phi',\zeta') \left(t^{(3)}_{\xi\zeta} + t^{(3)}_{\tau\xi}\right) \\
\A &&	+ \frac{2G\theta^3}{c^4}\partial_{\xi} \int_\alpha^\beta d\zeta' \int_0^{\rho_0(\zeta')} d\rho'\, \rho' \\
\A &&  \int_{0}^{2\pi}d\phi' \, \lim_{B\rightarrow \infty} K_B(\rho',\phi',\zeta')  \left(t^{(3)}_{\xi\zeta} + t^{(3)}_{\tau\xi}\right)\;.
\end{eqnarray}
The relevant combination of derivatives of the function $K_B$ with the approximation given in equation~(\ref{abd}) is given by
\begin{eqnarray}
	&& \partial_\chi \log\left(4\frac{B^2}{\rho''^2}\right) \left(t^{(3)}_{\xi\zeta} + t^{(3)}_{\tau\xi}\right) \\
&&\A - \partial_\xi \log\left(4\frac{B^2}{\rho''^2}\right)\left(t^{(3)}_{\chi\zeta} + t^{(3)}_{\tau\chi}\right) \\
\A &=& -\frac{P_0}{\pi c}|\mu'|^6 \rho'^2 e^{-2|\mu'|^2\rho'^2} \Big(\lambda \rho'\partial_{\rho'} + \theta\zeta'\partial_{\phi'}\Big) \log\big(\rho''^2\big)\,.
\end{eqnarray}
Again, the term containing the derivative with respect to $\phi'$ vanishes under the integration over $\phi'$ and we obtain
\begin{eqnarray}
\A && -\frac{P_0\lambda}{\pi c}|\mu'|^6\int_0^{\rho_0(\zeta')} d\rho'\, \rho'^3 \int_{0}^{2\pi}d\phi' \, \\
 && \A e^{-2|\mu'|^2\rho'^2} \rho'\partial_{\rho'} \log\big(\rho''^2\big)\\
\A &=& \frac{P_0\lambda}{c}|\mu'|^4\int_0^{\rho_0(\zeta')} d\rho'\, \Theta(\rho'-\rho) \rho'^2 \partial_{\rho'}e^{-2|\mu'|^2\rho'^2}\\
\A &=& \frac{P_0\lambda}{c}|\mu'|^4 \left\{\begin{array}{cc}
	\int_\rho^{\rho_0(\zeta')} d\rho'\, \rho'^2 \partial_{\rho'}e^{-2|\mu'|^2\rho'^2} &:\, \rho \le \rho_0(\zeta')\\
	0 &:\, \rho > \rho_0(\zeta')
	\end{array}\right\} \\
\A &=& -\frac{P_0\lambda}{c}|\mu'|^4  \Theta(\rho_0(\zeta') - \rho)\Bigg[2 \int_\rho^{\rho_0(\zeta')} d\rho'\, \rho' e^{-2|\mu'|^2\rho'^2} \\
\A && + \Big(\rho^2 e^{-2|\mu'|^2\rho^2} - \rho_0(\zeta')^2 e^{-2|\mu'|^2\rho_0(\zeta')^2}\Big) \Bigg] \\
\A &=& -\frac{P_0\lambda}{c}|\mu'|^2  \Theta(\rho_0(\zeta') - \rho)\Bigg[- \frac{1}{2}\int_\rho^{\rho_0(\zeta')} d\rho'\, \partial_{\rho'} e^{-2|\mu'|^2\rho'^2} \\
\A && + |\mu'|^2\Big(\rho^2 e^{-2|\mu'|^2\rho^2} - \rho_0(\zeta')^2 e^{-2|\mu'|^2\rho_0(\zeta')^2}\Big) \Bigg] \\
\A &=& -\frac{P_0\lambda}{2c}|\mu'|^2  \Theta(\rho_0(\zeta') - \rho) \Big((1 + 2|\mu'|^2 \rho^2) e^{-2|\mu'|^2\rho^2} \\
 &&  - (1 + 2|\mu'|^2 \rho_0(\zeta')^2) e^{-2|\mu'|^2\rho_0(\zeta')^2}\Big) \,.
\end{eqnarray}
Finally, the rotation of polarization for the parallel co-propagating light ray is given by
\begin{eqnarray}\label{blafin}
\Delta_{+} &=& \A { \lambda\frac{GP_0\theta^3}{c^5} \int_\alpha^\beta d\zeta'\,} \\
  \A  && \Theta(\rho_0 - |\mu'|\rho)|\mu'|^2 \Big((1 + 2|\mu'|^2 \rho^2) e^{-2|\mu'|^2\rho^2} \\
 &&  - (1 + 2\rho_0^2) e^{-2\rho_0^2}\Big)\;,
\end{eqnarray}
which leads to equation (\ref{bla}) for $\rho_0\rightarrow \infty$. In this case, we find that
\begin{equation}
	\Delta_{+} = -\frac{\theta^2}{8}\Big(1 - \partial_\sigma\Big)\Delta_{-}(\sqrt{\sigma}\rho)\Big|_{\sigma=1}\;.
\end{equation}
Again, we find that $\Delta_+$ vanishes if there is no overlap with the beam, i.e.~if $\rho > \rho_0(\alpha)$ and $\rho > \rho_0(\beta)$. For $\rho_0\rightarrow\infty$, $\rho \gg \theta\beta$ and $\rho \gg -\theta\alpha$, we find 
\begin{eqnarray}\label{eq:decaypl}
 \A \Delta_{+}
\A &=& \lambda\frac{GP_0\theta^3}{2c^5} \Bigg(\beta\Bigg(\frac{1}{\rho^2} + \frac{1}{(\theta\beta)^2}\Bigg) e^{-2(\rho/\theta\beta)^2} \\\A
&& -  \alpha\Bigg(\frac{1}{\rho^2} + \frac{1}{(\theta\alpha)^2}\Bigg) e^{-2(\rho/\theta\alpha)^2}\Bigg) \\
 &\approx &  \lambda\frac{GP_0\theta}{2c^5} \Bigg(\frac{1}{\beta} e^{-2(\rho/\theta\beta)^2}+ \frac{1}{|\alpha|} e^{-2(\rho/\theta\alpha)^2}\Bigg) \,.
\end{eqnarray}
For $\Delta_{t^\pm}$ for a finitely extended source beam and an infinitely extended test ray we obtain, considering only the leading order contribution,
 \begin{eqnarray}
\Delta^{(0)}_{t^\pm} & = &\nonumber \pm\frac{1}{2w_0^2}\int_{-\infty}^\infty d\xi \,\partial_\chi h_{\tau\zeta}^{(0)}\\
\A &=& \mp\frac{2G}{c^4} \int_\alpha^\beta d\zeta' \int_{\rho\le \rho_0(\zeta')}d\xi'd\chi'  \, \\
&& \lim_{B\rightarrow \infty} \partial_\chi K_{t,B}(\xi',\chi',\zeta') t_{\tau\zeta}^{(0)}
\,,
\end{eqnarray}
where the function $K_{t,B}$ is given by
\begin{align}
	&  K_{t,B}(\xi',\chi',\zeta') \\
	\A& = \log\left(\frac{B-\xi' + (\chi''^2 + (\zeta-\zeta')^2 + (B-\xi')^2)^{1/2}}{-B-\xi' + (\chi''^2 + (\zeta-\zeta')^2 + (B+\xi')^2)^{1/2}}\right)\,,
\end{align} 
and where $\chi'' = \chi' - \chi$.
For $B\gg 1$, we obtain
\begin{widetext}
\begin{eqnarray}
	\A&& K_{t,B}(\xi',\chi',\zeta') \\
	\A & = & \log\left(\frac{B-\xi' + (B-\xi')\Big(1 + (\chi''^2 + (\zeta-\zeta')^2)/(B-\xi')^2\Big)^{1/2}}{-B-\xi' + (B+\xi')\Big(1 + (\chi''^2 + (\zeta-\zeta')^2)/(B+\xi')^2\Big)^{1/2}}\right)\\
	\A & \approx & \log\left(\frac{B-\xi' + (B-\xi')\Big(1 + (\chi''^2 + (\zeta-\zeta')^2)/(2(B-\xi')^2)\Big)}{-B-\xi' + (B+\xi')\Big(1 + (\chi''^2 + (\zeta-\zeta')^2)/(2(B+\xi')^2)\Big)}\right)\\
	\A & = & \log\left(\frac{2(B-\xi') + (\chi''^2 + (\zeta-\zeta')^2)/(2(B-\xi'))}{(\chi''^2 + (\zeta-\zeta')^2)/(2(B+\xi'))}\right)\\
	\A & \approx & \log\left(\frac{B+\xi'}{B-\xi'} + \frac{4B^2}{\chi''^2 + (\zeta-\zeta')^2}(1-\xi'^2/B^2)\right)\\
	 & \approx & \log\left(\frac{4B^2}{\chi''^2 + (\zeta-\zeta')^2}\right)\,.
\end{eqnarray}
\end{widetext}
With the derivative of $K_{t,B}$ with respect to $\chi$,
\begin{equation}
	\partial_\chi  \log\left(\frac{4B^2}{\chi''^2 + (\zeta-\zeta')^2}\right) = 2\frac{\chi-\chi'}{\chi''^2 + (\zeta-\zeta')^2}
\end{equation}
we obtain for the zeroth order of the rotation of polarization of the transversal test ray
\begin{eqnarray}\label{eq:Deltat0}
\Delta^{(0)}_{t^\pm} & = & \mp \frac{4GP_0}{\pi c^5} \int_\alpha^\beta d\zeta' \int_{\rho\le \rho_0(\zeta')}d\xi'd\chi'  \\
\A && \frac{\chi-\chi'}{\chi''^2 + (\zeta-\zeta')^2}  |\mu'|^2 e^{- 2|\mu'|^2\rho'^2}
 \,.
\end{eqnarray}
{ Note that for $\chi=0$, the integrand is anti-symmetric in $\chi'$ and $\Delta^{(0)}_{t^\pm}$ vanishes.
For the first order contribution, we find}
\begin{eqnarray}
\Delta^{(1)}_{t^\pm}\A & = & \lambda\frac{4GP_0}{\pi c^5} \int_\alpha^\beta d\zeta' \int_{\rho\le \rho_0(\zeta')}d\xi'd\chi'  \\
 && \frac{\chi'(\chi-\chi')}{(\chi-\chi')^2 + (\zeta-\zeta')^2}  |\mu'|^4  e^{- 2|\mu'|^2\rho'^2}
\,.\label{hib}
\end{eqnarray}
{ For $\chi=0$, the integrand is symmetric in $\chi'$ and $\Delta^{(1)}_{t^\pm}$ does not vanish.}

%===========================================================
\section{Multipole expansion of the far field for finitely extended source and test beams\label{mult}}
For the finitely extended source beam, one can get analytical
approximations of $\Delta$ in the far field.  For simplicity we assume
here that the source beam extends from $-\beta$ to $\beta$, and the
probe beam from $-B$ to $B$.  The maximal radial extension of the
source beam, reached at $\zeta'=\pm \beta$, is then given by
$\rho'=\theta\beta/\sqrt{2}$.  This is the maximum scale on which all
components of 
the energy-stress tensor and its derivatives fall off like a Gaussian
(for smaller values of $|\zeta'|$ the decay is even faster).
Far field means then that the probe beam should be a distance $\rho\gg
\theta\beta/\sqrt{2}$ from the source beam when passing parallel to
the source beam. A much shorter distance of order $\rho\simeq 1$
suffices for the transversal beam passing at the beam waist for being
in the far field regime. \\
From eqs.(\ref{eq:primeli},\ref{pm}) we obtain, after shifting
derivatives to the prime-coordinates and partial integration,
\begin{align}
  \A \Delta_\pm=& -\frac{2G}{c^4\theta}\int_{-B}^B d(\theta \zeta)\int
               d^3x'\frac{1}{|\vec{x}-\vec{x}'|}\\
               & \Big[ \partial_{\chi'}(T_{\xi\zeta}(\vec{x}')\pm
               T_{\tau\xi}(\vec{x}'))\nonumber\\
& -\partial_{\xi'}(T_{\chi\zeta}(\vec{x}')\pm T_{\tau\chi}(\vec{x}'))\Big]\,.
\end{align}
For the partial integration we assume once more that we are in the
far-field, so that boundary terms are exponentially 
suppressed 
through the
Gaussian factor $\exp(-2|\mu|^2\rho^2)$. 
The source term relevant for $\Delta_-$ is
given to first order in $\theta$ by (see Appendix A,
eqs.\eqref{eq:38}
\begin{eqnarray}
  \label{eq:S-}
  S_-(\rho',\zeta')&\equiv& \frac{\pi
                            c}{4P_0\theta}\Big[\partial_{\chi'}(T_{\xi\zeta}(\vec{x}')
 -
               T_{\tau\xi}(\vec{x}')) \nonumber\\
&& -\partial_{\xi'}(T_{\chi\zeta}(\vec{x}')- T_{\tau\chi}(\vec{x}'))\Big]\nonumber\\
&=&\frac{2e^{-2\frac{\rho'^2}{1+\theta^2\zeta'^2}}\lambda(1+\theta^2\zeta'^2
  -2\rho'^2)}{(1+\theta^2\zeta'^2)^3}\,.
\end{eqnarray}
Manifestly, $S_-$ enjoys azimuthal symmetry.  It is then useful to
expand the function $1/|\vec{x}-\vec{x}'|$ as (see
e.g.~\cite{Jackson75} p. 93)
\begin{equation}
  \label{eq:1/r}
  \frac{1}{|\vec{x}-\vec{x}'|}=\sum_{l=0}^\infty \frac{r_<^l}{r_>^{l+1}}P_l(\cos\vartheta')P_l(\cos\vartheta)\,
\;,
\end{equation}
where $P_l$ are the Legendre-polynomials, $r_<$ ($r_>$) is the smaller (larger) of $|\vec{x}|$ and
$|\vec{x}'|$, and $\vartheta$ ($\vartheta'$) the angle between the $z$-axis
and $\vec{x}$ ($\vec{x}'$). For calculating the far field, we can set
everywhere 
$r_>=r=|\vec{x}|$ and $r_<=r'=|\vec{x}'|$.
This leads to {\small
\begin{eqnarray}
 \Delta_-&=& \sum_{l=0}^\infty \Delta_-^{(l)}\\
\A &=&-\frac{16 G P_0
           \theta}{c^5} \sum_{l=0}^\infty \int_{-B}^B d\zeta
           \frac{Q_-^{(l)}}{(\rho^2+\zeta^2)^{(l+1)/2}}P_l(\frac{\zeta}{\sqrt{\rho^2+\zeta^2}})\,,\label{D-}
\end{eqnarray}}
where the multipoles $Q_-^{(l)}$ are given by 
\begin{align}\A
  Q_-^{(l)}=&\int_{-\beta}^\beta d\zeta' \int_0^\infty \rho'
              d\rho'\,(\rho'^2+\zeta'^2)^{l/2}\\
&\times P_l(\frac{\zeta'}{\sqrt{\rho'^2+\zeta'^2}})S_-(\rho',\zeta')\,,
\end{align}
and we have used that in cylinder coordinates
$\vartheta=\arccos(\zeta/\sqrt{\rho^2+\zeta^2})$, and correspondingly for
$\vartheta'$. 
The multipoles and their contributions to $\Delta_-$ can be
calculated analytically.  All odd multipoles vanish, and so do the
monopole and dipole contribution ($l=0,1$, respectively).  $\Delta_-$
is then dominated by the quadropole contribution $l=2$.  The
correction due to higher
order multipoles $l=4,6,...$ decays quickly with $l$. We therefore
limit ourselves to listing the results for $l=2,4,6$.  Note that
the direct dependence on $\zeta'$ of $1/|\vec{x}-\vec{x}'|$ (rather
than on $\theta \zeta$ as for the rest of the integrand) brings about
additional $\theta$ dependence.  Neglecting these higher order terms,
we find
\begin{eqnarray}
  \label{eq:Q-}
  Q_-^{(2)}&=&\frac{\beta \lambda}{4}\;,\\
  Q_-^{(4)}&=&\frac{\beta \lambda}{8}(-3+4\beta^2)\;,\\
  Q_-^{(6)}&=&\frac{3\beta \lambda}{64}(15-40\beta^2+16\beta^4)\,,
\end{eqnarray}
and, with $\Omega\equiv 8\lambda \theta GP_0/c^5$, 
\begin{eqnarray}
  \label{eq:D-}
  \Delta_-^{(2)}&=&\frac{\Omega\beta}{2}\frac{B}{(B^2+\rho^2)^{3/2}}\;,\\
  \Delta_-^{(4)}&=&\frac{\Omega\beta(-3+4\beta^2)}{16} \frac{(2B^3-3B\rho^2)}{(B^2+\rho^2)^{7/2}}\;,\\ 
 \A \Delta_-^{(6)}&=&\frac{\Omega\beta(15-40\beta^2+16\beta^4)}{256}\\
  && \frac{(8B^4-40B^2\rho^2+15\rho^4)}{(B^2+\rho^2)^{11/2}}\,.
\end{eqnarray}

For $\Delta_+$, the lowest contributing terms are from the derivatives
of the third order of the metric. The expression for $S_-$ is replaced
by $S_+$ given by
\begin{eqnarray}
  \label{eq:S+}
 \A S_+(\rho',\zeta')&\equiv& \frac{\pi c}{P_0\theta^2}\Big[\partial_{\chi'}(T_{\xi\zeta}(\vec{x}')+
                            T_{\tau\xi}(\vec{x}'))\nonumber\\
                          &&  -\partial_{\xi'}(T_{\chi\zeta}(\vec{x}')+ T_{\tau\chi}(\vec{x}'))\Big]\\
        \A           &=&-\frac{
                       e^{-2\frac{\rho'^2}{1+\theta^2\zeta'^2}}\lambda\rho'^2(1-\rho'^2/(1+\theta^2\zeta'^2))}{(1+\theta^2\zeta'^2)^3}\,.
\end{eqnarray}
Also here the monopole contribution ($l=0$) and all contributions with
odd $l$, in particular the dipole contribution ($l=1$) vanish.  The
lowest order non-vanishing contributions are 
\begin{eqnarray}
  \label{eq:Q-}
  Q_+^{(2)}&=&-\frac{\beta \lambda \theta^2}{16}\;,\\
  Q_+^{(4)}&=&\frac{\beta \lambda \theta^2}{64}(9-8\beta^2)\;,\\
  Q_+^{(6)}&=&-\frac{3\beta \lambda\theta^2}{128}(15-30\beta^2+8\beta^4)\,,
\end{eqnarray}
to be substituted into the expression corresponding to \eqref{D-}, i.e. 
{\small\begin{eqnarray}
 \Delta_+&=& \sum_{l=0}^\infty \Delta_+^{(l)}\\
\A &=&-\frac{16 G P_0
           \theta}{c^5} \sum_{l=0}^\infty \int_{-B}^B d\zeta
           \frac{Q_+^{(l)}}{(\rho^2+\zeta^2)^{(l+1)/2}}P_l(\frac{\zeta}{\sqrt{\rho^2+\zeta^2}})\,.\label{D+}
\end{eqnarray}}
This leads to 
\begin{eqnarray}
  \label{eq:D-}
  \Delta_+^{(2)}&=&-\frac{\Omega\theta^2}{8}\frac{B\beta}{(B^2+\rho^2)^{3/2}}\;,\\
  \Delta_+^{(4)}&=&\frac{\Omega\theta^2(-9+8\beta^2)}{128}\frac{(-2B^3+3B\rho^2)}{(B^2+\rho^2)^{7/2}}\;,\\ 
 \A \Delta_+^{(6)}&=&-\frac{\Omega\theta^2 B\beta(15-30\beta^2+8\beta^4)}{512}\\
  && \frac{(8B^4-40B^2\rho^2+15\rho^4)}{(B^2+\rho^2)^{11/2}}\,,
\end{eqnarray}
where we recall that $\Omega$ contains already one factor $\theta$.
So both $\Delta_\pm$ fall off as $1/\rho^3$ in the far-field due to the
quadrupole contribution.  For fixed $\rho,\beta$ that contribution
decays as $1/B$ for large $B$, i.e.~$B\gg \rho$.
This can be traced back to the integral over $\zeta$ in 
 and would not be the case for the monopole
 contribution.

For $\Delta_{t^\pm}$ we start with the lowest, zeroth order
in $\theta$. It is then useful to keep the derivatives of the
energy-stress tensor outside the calculation of the multipoles, as
otherwise the cylindrical symmetry gets spoiled.  We find
\begin{align}
  \Delta^{(0)}_{t^\pm}=&\mp \frac{8GP_0}{
                   c^5}\int_{-B}^B d\xi \, \partial_\chi\sum_{l=0}^\infty
                   \frac{P_l(\frac{\zeta}{\sqrt{\rho^2+\zeta^2}})}{(\rho^2+\zeta^2)^{(l+1)/2}}Q_{t^\pm}^{(0)(l)}\;,\\
Q_{t^\pm}^{(0)(l)}=&\int_{-\beta}^\beta{ d\zeta'}\int_0^\infty d\rho'
  \rho'(\rho'^2+\zeta'^2)^{l/2}P_l(\frac{\zeta'}{\sqrt{\rho'^2+\zeta'^2}})\nonumber
\\&\times\frac{1}{1+\theta^2\zeta'^2}e^{-{ 2}\frac{\rho'^2}{1+\theta^2\zeta'^2}}\;. 
\end{align}
Also here, all the odd-power multipoles $(l=1,3,5,\ldots)$ vanish due to
the fact that the Legendre-polynomials of odd order are odd, whereas
the rest of the integrand in $Q_{t^\pm}^{(0)(l)}$ is even in $\zeta'$. 
The three lowest non-vanishing
multipoles read
\begin{align}
  Q_{t^\pm}^{(0)(0)}=&\frac{\beta}{2}\;,\\
  Q_{t^\pm}^{(0)(2)}=&-\frac{1}{24}\beta(3-4\beta^2)\;,\\
  Q_{t^\pm}^{(0)(4)}=&\frac{1}{160}\beta(15-40\beta^2+16\beta^4) \;.
\end{align}
The corresponding contributions to $\Delta_{t^\pm}$ at $\zeta=0$ are
\begin{eqnarray}
  \Delta_{t^\pm}^{(0)(0)}&=&\pm\tilde{\Omega}\frac{B\beta}{\chi\sqrt{B^2+\chi^2}}\;,\\
  \Delta_{t^\pm}^{(0)(2)}&=&\pm\tilde{\Omega}\frac{B\beta(3-4\beta^2)(2B^2+3\chi^2) }{24\chi^3(B^2+\chi^2)^{3/2}}\;,\\
  \A \Delta_{t^\pm}^{(0)(4)}&=&\pm\tilde{\Omega}\frac{B\beta(15-40\beta^2+16\beta^4)}{640\chi^5(B^2+\chi^2)^{5/2}} \\
  && (8B^4+20B^2\chi^2 + 15\chi^4) \,,
\end{eqnarray}
where $\tilde{\Omega}=8GP_0/c^5$. 
We see that now there is a
contribution from the monopole that leads to a decay as $1/\chi^2$
with the minimal distance $\chi$ from the beamline when evaluated at
$\zeta=0$ and in the limit of 
$\chi\gg B$. The next (quadrupole) term contributes a $1/\chi^4$
decay. In the limit of $B\to\infty$ at fixed $\chi$, the monopole
contribution converges to a $\beta/\chi^2$ behavior.   \\ 
  
For the first order term in $\Delta_{t^\pm}$, the contribution to the
Faraday effect, we obtain with the expressions for the energy-momentum tensor given in appendix~\ref{A}, using the symmetry of $|\vec x-\vec x'|$ and performing a partial integration,
\begin{eqnarray}\A
\Delta^{(1)}_{t^\pm}
\A &=& \frac{2G\theta}{c^4}\partial_{\chi} \int_{-B}^B d\xi \int_{-\infty}^\infty d\xi' d\chi' d\zeta' \, \frac{1}{|\vec{x}-\vec{x}'|} t^{(1)}_{\xi\zeta}\\
\A &=& \frac{\lambda\theta}{4} \partial_{\chi} \Delta^{(0)}_{t^+}  -\frac{GP_0\theta^2}{\pi c^5}\partial_{\chi} \int_{-B}^B d\xi \,\partial_\xi \int_{-\infty}^\infty d\xi' d\chi'  \, \\
 &&\int_{-\beta}^\beta d\zeta' \frac{1}{|\vec{x}-\vec{x}'|}\frac{\zeta'}{1+\theta^2\zeta'^2}e^{-2\frac{\rho'^2}{1+\theta^2\zeta'^2}} \,.
\end{eqnarray}
We neglect the second term as it is of higher order in $\theta$. For the first term, we find
from the multipole expansion of $\Delta^{(0)}_{t^+}$ for $\zeta=0$
\begin{eqnarray}
  \Delta_{t^\pm}^{{(1)}(0)}&=&-\frac{\lambda\theta}{4}\tilde{\Omega}\frac{B\beta(B^2+2\chi^2)}{\chi^2(B^2+\chi^2)^{3/2}}\;,\\
 \A \Delta_{t^\pm}^{{(1)}(2)}&=&-\frac{\lambda\theta}{4}\tilde{\Omega}\frac{B\beta(3-4\beta^2)}{8\chi^4(B^2+\chi^2)^{5/2}}\\
  && (2B^4+5B^2\chi^2+4\chi^4)\;,\\
  \A \Delta_{t^\pm}^{{(1)}(4)}&=&-\frac{\lambda\theta}{4}\tilde{\Omega}\frac{B\beta(15-40\beta^2+16\beta^4)}{128\chi^6(B^2+\chi^2)^{7/2}} \\
  && (8B^6 + 28B^4\chi^2 + 35B^2\chi^4 + 18\chi^6) \,.
\end{eqnarray}  
In a real experiment, it should be kept in mind that the gravitational
effects from emitter and absorber and the 
power-supplies feeding them, as well as heat-radiation from the
absorber may lead to effects that mask the rotation of the polarization of the
source beam itself in the far
field, if their dipole- or monopole-contributions do not vanish.  If
one wishes to evaluate these effects, a careful modelling of the entire
setup will be necessary.

%===========================================================
\section{The infinitely thin beam\label{G}}

The metric perturbation induced by an infinitely thin beam of light that
extends along the $\zeta$-axis from $-\beta$ to $\beta$ is given by the only
non-zero components $h_{\tau\tau} = -h_{\tau\zeta} = h_{\zeta\zeta} = h$,
where $h$ is given as \cite{tolman}
\begin{equation}
	h = \frac{4GP_0w_0^2}{c^5} \log\left(\frac{\beta - \zeta + (\rho^2 + (\beta-\zeta)^2)^{1/2}}{-\beta - \zeta + (\rho^2 + (\beta+\zeta)^2)^{1/2}}\right)\,.
\end{equation}
Therefore, we find with equation (\ref{eq:rotpleb}) at $\zeta=0$ and for large $\chi$
\begin{eqnarray}
	\A \Delta_{t^\pm} & \approx & \pm\frac{1}{2w_0^2}\int_{-B}^B d\xi \,\partial_\chi h_{\tau\zeta}^{(0)} \\
	&\approx & \pm \frac{8GP_0}{c^5} \frac{\beta B}{\chi\sqrt{B^2+\chi^2}}\,,
\end{eqnarray}
where we considered a test ray extending from $-B$ to $B$, and
\begin{eqnarray}
	\Delta_{t^\pm} & \approx \pm \frac{8GP_0}{c^5} \frac{\beta}{\chi}\,,
\end{eqnarray}
for the infinitely extended test ray.

%======================================================================================
\bibliographystyle{unsrt}
\bibliography{biblio}

%======================================================================================
\end{document}